\documentclass[12pt,preprint]{aastex}
\usepackage{graphicx}
\usepackage{natbib}

\begin{document}

\title{Observations and Modelling of Relativistic Spin Precession in PSR J1141$-$6545}

\author{R. N. Manchester\altaffilmark{1}, M. Kramer\altaffilmark{2,3},
  I. H. Stairs\altaffilmark{4,1,5}, M. Burgay\altaffilmark{6},
  F. Camilo\altaffilmark{7}, G. B. Hobbs\altaffilmark{1},
  D. R. Lorimer\altaffilmark{8}, A. G. Lyne\altaffilmark{2},
  M. A. McLaughlin\altaffilmark{8}, C. A. McPhee\altaffilmark{4},
  A. Possenti\altaffilmark{6}, J. E. Reynolds\altaffilmark{1}, W. van
  Straten\altaffilmark{5}}

\altaffiltext{1}{CSIRO Astronomy and Space Science, Australia
  Telescope National Facility, PO Box 76, Epping NSW 1710, Australia
  (dick.manchester@csiro.au)} \altaffiltext{2}{Jodrell Bank Centre for
  Astrophysics, University of Manchester, Manchester M13 9PL, UK}
\altaffiltext{3}{Max-Planck-Institut f\"ur Radioastronomie, Auf dem
  H\"ugel 69, 53121 Bonn, Germany} \altaffiltext{4}{Department of
  Physics and Astronomy, University of British Columbia, 6224
  Agricultural Road, Vancouver, B.C. V6T 1Z1, Canada}
\altaffiltext{5}{Centre for Astrophysics and Supercomputing, Swinburne
  University of Technology, PO Box 218, Hawthorn VIC 3122, Australia}
\altaffiltext{6}{INAF-Osservatorio Astronomico di Cagliari,
  Loc. Poggio dei Pini, 09012 Capoterra (CA), Italy}
\altaffiltext{7}{Columbia Astrophysics Laboratory, Columbia
  University, New York, NY 10027} \altaffiltext{8}{Department of
  Physics, West Virginia University, Morgantown, WV 26506}

\begin{abstract}
  Observations of the binary pulsar PSR J1141$-$6545 using the Parkes
  radio telescope over 9.3 years show clear time-variations in pulse
  width, shape and polarization. We interpret these variations in
  terms of relativistic precession of the pulsar spin axis about the
  total angular momentum vector of the system changing our view of the
  emission beam. Over the nine years, the pulse width at the 50\%
  level has changed by more than a factor of three, reaching a maximum
  value of nearly $13\degr$ in early 2007. Large variations have also
  been observed in the 1400-MHz mean flux density; this reached a peak
  of $\sim 20$~mJy in mid-2002 but over the past several years has
  been relatively steady at $\sim 3$~mJy. The pulse polarization has
  been monitored since 2004 April using digital filterbank systems and
  also shows large and systematic variations in both linear and
  circular polarization. Position angle variations, both across the
  pulse profile and over the data span, are complex, with major
  differences between the central and outer parts of the pulse
  profile. We interpret the outer parts as representing the underlying
  magnetic field and fit the rotating-vector model to these regions.
  Modelling of the observed position angle variations by relativistic
  precession of the pulsar spin axis shows that the spin-orbit misalignment
  angle is about $110\degr$ and that the precessional phase has passed
  through $180\degr$ during the course of our observations. At the
  start of our observations, the line-of-sight impact parameter was
  about $4\degr$ in magnitude and it reached a minimum very close to
  $0\degr$ around early 2007, consistent with the observed pulse width
  variations. We have therefore mapped approximately one half of the
  emission beam, at least out to a radius of about $4\degr$, showing
  that it is very asymmetric with respect to the magnetic axis. The
  derived precessional parameters imply that the pre-supernova star
  had a mass of about 2~M$_\odot$ and that the supernova recoil kick
  velocity was relatively small, between 100 and 250~km~s$^{-1}$,
  depending on the assumed systemic velocity. With the reversal in the
  rate of change of the impact parameter, we predict that over the
  next decade we will see a reversed ``replay'' of the variations
  observed in the past decade.
\end{abstract}


\section{Introduction}\label{sec:intro}

Binary pulsars with short orbital periods and massive companions
exhibit a range of relativistic effects, the most prominent of which
are relativistic periastron precession, gravitational time dilation,
transverse Doppler effects, orbit decay due to loss of energy to
gravitational waves and the Shapiro delay resulting from the passage of
the pulsar signal through the gravitational field of the
companion. These effects may be described by the so-called
``Post-Keplerian'' parameters \citep{dd85,dd86} and are most readily
observed in double-neutron-star systems such as the Hulse-Taylor
binary pulsar, PSR B1913+16 \citep{wt05} and the Double Pulsar, PSR
J0737$-$3039A/B \citep{ksm+06}. Another relativistic effect which is
potentially observable in such systems is the precession of
the pulsar spin axis resulting from coupling between the spin and
orbital angular momenta \citep{dr74,bo75}. The angular rate of
precession, $\Omega_p$, is given by:
\begin{equation}
\Omega_p = \frac{G^{2/3}}{c^2} \left(\frac{P_b}{2\pi}\right)^{-5/3}
  \frac{m_c(4m_p + 3m_c)}{(1-e^2)(m_p + m_c)^{4/3}}
\label{eq:prec_freq}
\end{equation}
where $G$ is the gravitational constant, $c$ is the velocity of light,
$P_b$ is the binary orbital period, $e$ is the orbital eccentricity
and $m_p$ and $m_c$ are the pulsar and companion masses
respectively. For PSR B1913+16, the expected rate from
Equation~(\ref{eq:prec_freq}) is $1\fdg21$ yr$^{-1}$, corresponding to a
precessional period of about 300 years, whereas for PSR
J0737$-$3039A/B the rate is about a factor of four
larger and the precessional period is about 75 years. 

Since the emission from pulsars is believed to be beamed, probably
along the open field lines associated with the magnetic poles on the
neutron star, precession of the pulsar spin axis will result in
changes in the beam aspect as viewed from the Earth. One would expect
this to lead to changes in the observed pulse profile and changes
attributed to this effect have indeed been observed in PSR B1913+16
\citep{wrt89,kra98} and PSR B1534+12 \citep{arz95,sta04}. For PSR
B1913+16, a significant fraction of the expected beam diameter has
been traversed in the time since its discovery and this has been used
to map the two-dimensional structure of the emission beam
\citep{wt02,cw08} . Because of its relatively short precessional
period, similar effects were expected to be observable for the Double
Pulsar, but surprisingly, they have not so far been observed in the A
pulsar \citep{mkp+05,fsk+08}. A possible reason for this is that A's
spin-orbit misalignment angle (the angle between the spin and orbital angular
momenta) is small \citep[see also][]{wkh04,std+06}. However, large
variations in the pulse profile and orbital visibility of pulsar B
have been observed \citep{bpm+05} and these are certainly due to some
combination of relativistic periastron precession and precession of
the B spin axis. There is now direct evidence for the relativistic
spin precession of pulsar B. The emission from pulsar A is eclipsed
for about 30 seconds when it passes behind pulsar B \citep{lbk+04,krb+04}
with the eclipse profile showing modulation related to the spin period of
pulsar B \citep{mll+04}. Modelling of the long-term variations in the
eclipse profile are consistent with the relativistic precession of B's
spin axis with a spin-orbit misalignment angle of about $130\degr$ \citep{bkk+08}.

PSR J1141$-$6545 is a 394-ms pulsar in an eccentric 4.7-h binary orbit
with a relatively massive companion ($m_c \sim 1.0$~M$_\sun$)
discovered in the Parkes Multibeam Pulsar Survey \citep{klm+00a}. With
these characteristics, relativistic effects are expected to be
detectable and indeed, measurement of the relativistic advance of
periastron was reported in the discovery paper. The expected rate of
relativistic precession of the pulsar spin axis is 1\fdg36 yr$^{-1}$,
corresponding to a precessional period of about 265 yr. Unusually for
such a binary system, the pulsar characteristic age is relatively low,
about $1.4\times 10^6$ yr, suggesting that the initially more massive
primary star evolved to form a white dwarf, the present companion
star, and in the process transferred mass to the secondary which
subsequently exploded leaving behind a neutron star, the present
pulsar \citep{dc87,klm+00a,ts00a}. Continued timing observations
\citep{bokh03,bbv08} reinforced these ideas and gave detections of
relativistic time dilation and orbital decay which were consistent
with the predictions of general relativity. A clear orbital modulation
in the timescale for interstellar scintillation was detected by
\citet{obv02}, giving estimates of the orbital inclination angle $i$
(or $180-i$) $= 76\fdg0 \pm 2\fdg5$ and the system transverse space
velocity, $115 \pm 10$~km~s$^{-1}$. From measurements of the
relativistic Shapiro delay, \citet{bbv08} derive a value for the
inclination angle of $i=73\fdg0 \pm 1\fdg5$, consistent
with the scintillation measurement.

Observations at frequencies around 1.4 GHz by \citet{hbo05} showed
that the mean pulse profile for PSR J1141$-$6545 changed significantly
over the five years from 1999 July to 2004 May (MJD range 51381 --
53134), with an approximately linear increase in $W_{10}$, the pulse
width at 10\% of the pulse peak. Furthermore, based on two measurments
separated by about 0.7 yr, there appeared to be steepening of the
gradient of polarization position angle (PA) $d\psi / d\phi$, where
$\psi$ is the PA and $\phi$ is the pulse phase (measured in the same
units as the PA) near the profile center, at a rate of $2.3 \pm 0.4$
yr$^{-1}$. Variations of PA across pulse profiles are often well
described by the ``rotating-vector model'' (RVM) in which the emission
is assumed to be polarized parallel (or perpendicular) to the projected
direction of magnetic fields in the vicinity of a magnetic pole on the
rotating neutron star \citep{rc69a}. In
the RVM, $\psi$ varies as
\begin{equation}\label{eq:rvm}
\tan(\psi - \psi_0) = \frac{\sin\alpha \sin(\phi -
  \phi_0)}{\sin\zeta \cos\alpha - \cos\zeta
  \sin\alpha \cos(\phi - \phi_0)}
\end{equation}
$\psi_0$ is the PA at $\phi_0$, the center of symmetry of the PA
variation and the projected direction of the rotation axis for a
dipole field, $\zeta = \alpha + \beta$ is the inclination of the
observer's line of sight relative to the rotation axis of the star,
$\alpha$ is the inclination of the magnetic axis relative to the
rotation axis and $\beta$ is the impact parameter of the observer's
line of sight, that is, the minimum angle between the magnetic axis
and the observer's line of sight which occurs at $\phi =
\phi_0$.\footnote{Note that here and in Sections~\ref{sec:model},
  \ref{sec:beam} and \ref{sec:kick}, $\psi$ is defined in a
  right-handed coordinate system, increasing in the clockwise
  direction from East (looking toward the source) following
  \citet{dt92} (hereafter DT92) and \citet{ew01}. This is opposite to
  the IAU convention in which PA increases in the counterclockwise
  direction from North toward East looking toward the source. The
  polarization conventions are discussed in more detail in
  Section~\ref{sec:obs}.}  From Equation~\ref{eq:rvm},
\begin{equation}\label{eq:pa_slope}
(d\psi / d\phi)_{\rm m} = \sin\alpha/\sin\beta
\end{equation}
where $(d\psi / d\phi)_{\rm m}$ is the maximum value of $d\psi / d\phi$ at
$\phi_0$. Therefore, the increase in PA gradient observed by
\citet{hbo05} for PSR J1141$-$6545 was interpreted as a decrease in
$|\beta|$, implying that our line of sight was moving closer to the beam
axis. This and the changing pulse width were interpreted in terms of
relativistic precession of the pulsar spin axis with a spin-orbit
misalignment angle greater than $15\degr$ and probably less than $30\degr$.

In this paper we report on observations of PSR J1141$-$6545 made over
9.3 years from 1999 August to 2008 November with the Parkes radio
telescope at frequencies close to 1.4 GHz. Since 2004 April, we have
observed with systems recording full polarization data but before
that, only total intensity data were obtained.  The observations and
analysis procedures are described in Section~\ref{sec:obs} and results are
presented in Section~\ref{sec:results}. In Section~\ref{sec:model} we describe the
interpretation of the observed variations in terms of
relativistic precession of the pulsar spin axis and the implied shape of
the emission beam is described in Section~\ref{sec:beam}. The implications of
our precessional model for the formation of the system are described in
Section~\ref{sec:kick}. In Section~\ref{sec:summary} we summarize results and give
our conclusions.

\section{Observations and Analysis Procedures}\label{sec:obs}

We have observed PSR J1141$-$6545 using the Parkes 64-m radio
telescope between 1999 August and 2008 November (MJDs 51411 to 54785)
using (at different times) two receiving systems, either the center
beam of the Parkes 20-cm multibeam (MB) receiver \citep{swb+96} or the
``H-OH'' receiver. The MB receiver has a bandwidth of about 300 MHz
centered at about 1.4 GHz and a system equivalent flux density
($S_{\rm sys}$) of approximately 29 Jy. The H-OH receiver has a wider
bandwidth, approximately 600 MHz from 1.2 to 1.8 GHz, but a somewhat
higher $S_{\rm sys}$ of about 35 Jy.  Both receivers have orthogonal
linearly-polarized feeds and a linearly-polarized, broad-band and
pulsed calibration signal which can be injected into the feed at
$45\degr$ to the two signal probes. Four different backend systems
were used, all of which recorded data across a total bandwidth of 256
MHz: an analogue filterbank (AFB) having 512 frequency channels on
each polarization and recording total-intensity data using a one-bit
digitiser system \citep[see][]{mlc+01}, a wideband correlator (WBC)
which recorded all four Stokes parameters with 1024 frequency
channels, and two digital filterbank systems (PDFB1 and PDFB3) which
also recorded full polarization data with 512 channels across the
band. For the WBC and the PDFBs, each observation was preceded by a
short (2-min) observation of the pulsed calibration
signal. Observations of Hydra A, assumed to have a flux density of
43.1 Jy at 1.4 GHz, were used to set the flux-density scale. Off-line
processing made use of the {\sc psrchive} pulsar data analysis system
\citep{hvm04}.

A log of the observations is given in Table~\ref{tb:obs}. AFB
observations, which commenced soon after the discovery of the pulsar,
were recorded on tape and processed off-line to form mean total
intensity profiles. Frequency channels containing known interference
were given zero weight before data were summed across the band. For
the purposes of this paper, observations made over intervals ranging
from a few hours to a few weeks have been grouped as shown in
Table~\ref{tb:obs}. Pulse widths and flux densities were determined
for each observation and mean values and their uncertainties for the
group determined from these. The flux-density scale for the AFB data
was established by comparison with contemporaneous WBC observations.

WBC and PDFB observations were typically of 1-h duration and were
folded on-line to form 1-min sub-integrations.  Frequency channels
containing strong interference, affected by resonances in the feed or
with low system gain (normally 5\% of the band at the band edges) and
sub-integrations affected by strong impulsive interference were given
zero weight. Using the pulsed calibration signal as a reference, the
variations in instrumental gain and phase across the band were
removed, Stokes parameters formed and the data placed on a
flux-density scale. Data were summed to form 15-min sub-integrations
and mean profile widths and flux densities computed as for the AFB
data. All profiles were recorded with 1024 bins across the pulse
period.

Observed PAs given in this paper follow the astronomical convention,
with PA measured from celestial North and increasing toward East
(counter-clockwise looking toward the source). The IEEE definition of
circular polarization is adopted (i.e., at a given point in space and
looking at the source, the E-vector for a right-circular wave rotates
in a counter-clockwise direction). Both of these definitions are in
accord with IAU recommendations (Trans. IAU, 15B, 166, 1973). Stokes
$V$ is defined in the sense left-circular minus right-circular to
conform with established pulsar polarization conventions which pre-date
the IAU recommendations. File header
parameters defined the polarization setup for each receiver/back-end
combination and these were used by {\sc psrchive} to correct the
Stokes parameters so that they conformed to these definitions. For
further details on polarization conventions and their implementation,
see \citet{vmjr09}. The strong millisecond pulsar, PSR J0437$-$4715,
was observed almost every session and used to check the sign of the
calibrated Stokes parameters and the absolute value of position angle
by comparison with the results given by \citet{nms+97} which are known
to be in accordance with these definitions. The linearly polarized
intensity $L=(Q^2 + U^2)^{1/2}$ is a positive definite quantity which
is biased by noise. Plotted values have been corrected for this bias
using the relations given by \citet{lk05}.

Before summing in frequency, the observed pulse profiles must be
corrected for Faraday rotation across the band. The nominal rotation
measure (RM) for PSR J1141$-$6545 is $-84\pm 2$ rad m$^{-2}$
\citep{hml+06}. RMs were measured for each epoch having polarization
data by summing the upper and lower halves of the bandpass separately
using the nominal RM, taking a weighted mean of the PA differences
between the upper and lower bands across the profile, recomputing the
RM and then iterating until convergence. We show below that the
observed PA variations across the pulse profile are complex, both in
time and in frequency, with quite different behavior in the central
and outer parts of the profile. Consequently, RMs were determined
separately for the central and outer parts; we believe that the outer
parts represent the true interstellar RM.

\begin{figure*}
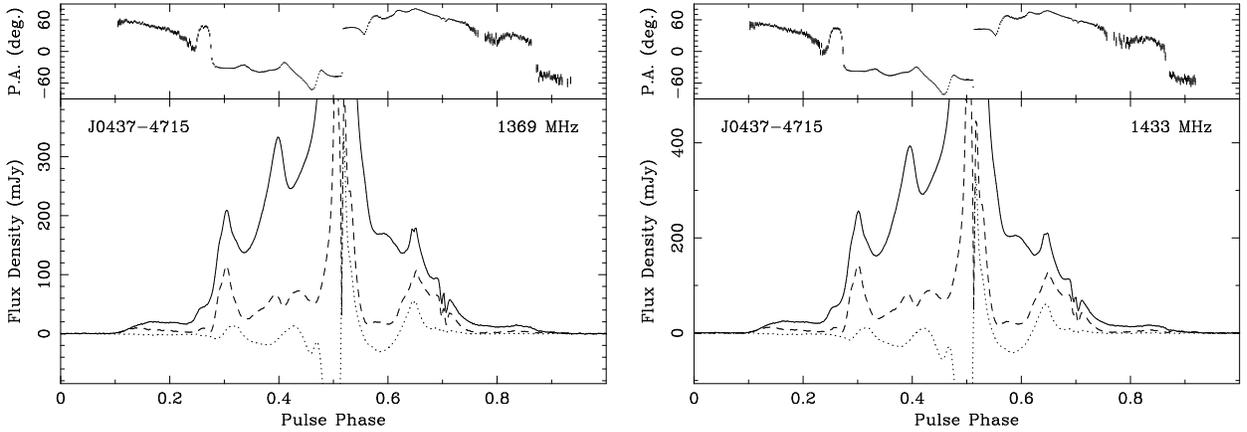

\begin{center} 
\begin{tabular}{cc}
\mbox{\includegraphics[angle=-90,width=80mm]{J0437_MB.ps}} &
\mbox{\includegraphics[angle=-90,width=80mm]{J0437_HOH.ps}} 
\end{tabular}
\caption{Mean pulse profiles and polarization parameters for PSR
  J0437$-$4715 at two epochs, 2006 October 03 (MJD 54011) using the
  multibeam receiver (left) and 2007 January 18 (MJD 54118) using the
  H-OH receiver (right). The respective integration times were 3.19~h
  and 2.13~h and the observing and data analysis methods were
  identical to those used for observations of PSR J1141$-$6545.  In
  the lower panel of each plot the solid line is the total intensity
  (Stokes $I$), the dashed line is the linearly polarized intensity
  and the dotted line is Stokes V. To better illustrate the details of
  the polarization variation, the plots have been truncated at 0.15 of
  the peak $I$ amplitude. The position angle of the linearly polarized
  component is shown in the upper panel of each plot. Note that
  postion angles are defined according to the IAU convention; they are
  absolute and apply to the frequency marked on each
  plot.}\label{fg:0437}
\end{center}
\end{figure*}

The MB receiver suffers from significant coupling between the
nominally orthogonal feed probes. Observed polarization variations
resulting from this coupling are a function of parallactic angle and
the coupling parameters can be measured by analysis of a series of
observations of a polarized source covering a wide range of
parallactic angles \citep{vs04}. PSR J0437$-$4715 was observed every
few months, typically for 10 minutes each hour during a 10-hour
transit, for each of the WBC and PDFB configurations used with the MB
receiver (Table~\ref{tb:obs}).  These data were analysed using the
{\sc psrchive} program {\sc pcm} to determine the feed ellipticities
and their relative orientation as a function of frequency across the
band. Observations of PSR J1141$-$6545 were then calibrated to remove
the effects of the cross-coupling. Cross-coupling in the H-OH receiver
is an order of magnitude less and this calibration step was not
required. Figure~\ref{fg:0437} shows (truncated) PDFB1 polarization
profiles for PSR J0437$-$4715 taken with the MB and H-OH receivers
using the observational and processing methods described above. The
results are essentially identical within the noise uncertainties,
confirming that our calibration procedures are robust.\footnote{The RM
  for PSR J0437$-$4715 is only +1.5 rad m$^{-2}$ \citep{nms+97} and so
  the PA difference expected between 1369 MHz and 1433 MHz is only
  about $0\fdg4$, not visible in Figure~\ref{fg:0437}.}

\section{Results}\label{sec:results}

While not the primary objective of this work, it was necessary to
monitor period variations of the pulsar in order make accurate
predictions of the topocentric period for use during the observations
and to sum observations in off-line processing. Initially, the timing
model of \citet{bokh03} was used for this purpose.  However, while
processing the 2007 July 18 observations, it was realised that the
pulsar had suffered a sizable glitch. Glitch parameters resulting from
a fit of a model including post-glitch exponential decay \citep[see,
e.g.][]{wmp+00} to pulse time-of-arrival (ToA) data from 2006 April to
2008 November are given in Table~\ref{tb:glt}.  ToAs were obtained
from average pulse profiles for data segments of 15-min duration and
the program {\sc tempo2} \citep{hem06} was used to determine the
timing parameters. There was only a small relaxation of the pulse
frequency toward the extrapolated pre-glitch solution following the
glitch, i.e., the $Q$ parameter was small. Significant timing noise
remains after the fit and the uncertainties have been multiplied by
five ($\sim \sqrt(\chi^2_r)$) to allow for this.

Table~\ref{tb:psrpar} gives pulsar timing parameters from a fit to
post-glitch data from 2007 December to 2008 November; the $\sim 160$~d
following the glitch were omitted to avoid biasing the result by the
post-glitch relaxation. Over the fitted data span there is significant
timing noise and the fit uncertainties have been multiplied by a
factor of three. Only the pulse frequency $\nu$ and $\dot\nu$ were
fitted for; the position and the binary parameters were held fixed at
the values given by \citet{bbv08}. Parameters are quoted in
TDB-compatible units, and the ``DD'' binary model \citep{dd86} and the Jet Propulsion
Laboratory Solar-System ephemeris DE405 \citep{sta98b} were used.

Figure~\ref{fg:w50s14} shows the observed time variations of the 50\%
mean pulse width ($W_{50}$) and mean flux density at 1400 MHz over the
9.3-year data span.  Such dramatic long-term variations in pulse width
(more than a factor of three) are unprecedented. The most
straight-foward explanation is that they result from precession of the
pulsar spin axis changing our view of the pulsar beam. The variations
are complex with a $\sim 400$-d interval around MJD 52800 (2003 June)
where the width clearly decreases with time before increasing
again. Recent data show that a maximum width was reached around MJD
54150 (2007 February) and now the width is decreasing. There is good
agreement between the widths obtained with the different back-end
systems, although the AFB widths tend to be a few percent less than those
measured with the WBC or PDFB1 at the same or similar time owing to
differences in the instrumental impulse response. The time variation
in $W_{10}$ is similar, but with a smaller relative change over the
data span.

The mean pulsed flux density has also changed dramatically over the
data span with a broad peak around MJD 52500 (mid-2002). There is a
significant day-to-day variation in the measured flux densities. This
is indicated by the large error bars on the AFB averages (which
typically cover a week or more) and by the scatter in the WBC/PDFB
measurements (which typically are made on a single day), both much
larger than the uncertainty in an individual measurement. These
short-term variations can be attributed to refractive scintillation
for which the expected timescale is about seven days
\citep{obv02}. However, the broad rise and decay of the flux density
over the nine-year data span is unlikely to be a scintillation effect
and we attribute this to the changing aspect of a complex beam pattern
sweeping across the Earth as the pulsar spin axis precesses.

\begin{figure}
\includegraphics[width=90mm]{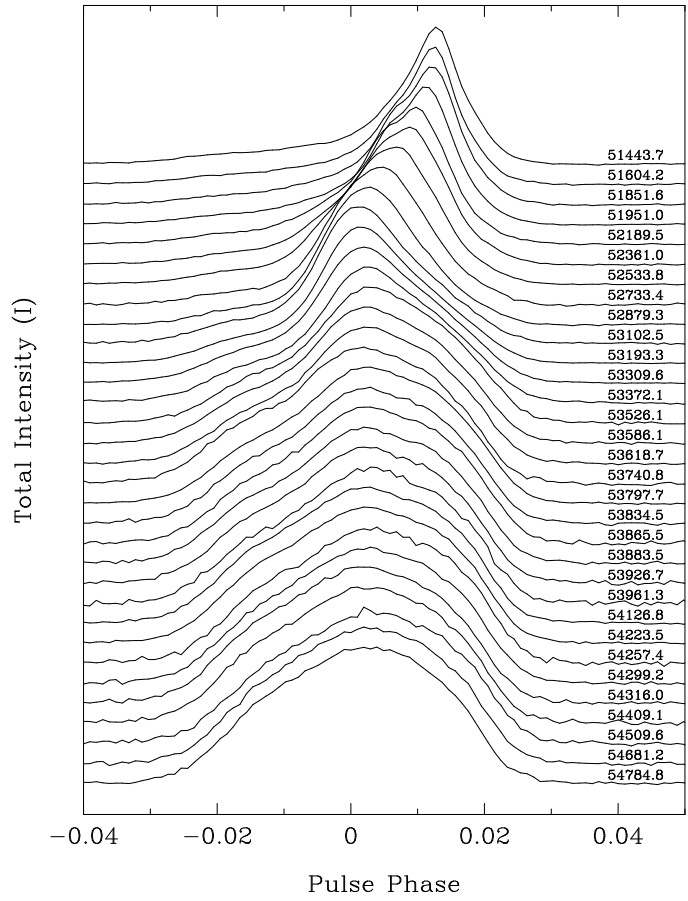}
\caption{Mean total intensity (Stokes $I$) pulse profiles for PSR
  J1141$-$6545 normalised to the same peak amplitude. The mean MJD for
  each profile is shown (note that time increases downward). The
  procedure for phase alignment of profiles is described in the text. }\label{fg:prf}
\end{figure}

\begin{figure*}
\begin{center} 
\begin{tabular}{cc}
\mbox{\includegraphics[width=75mm]{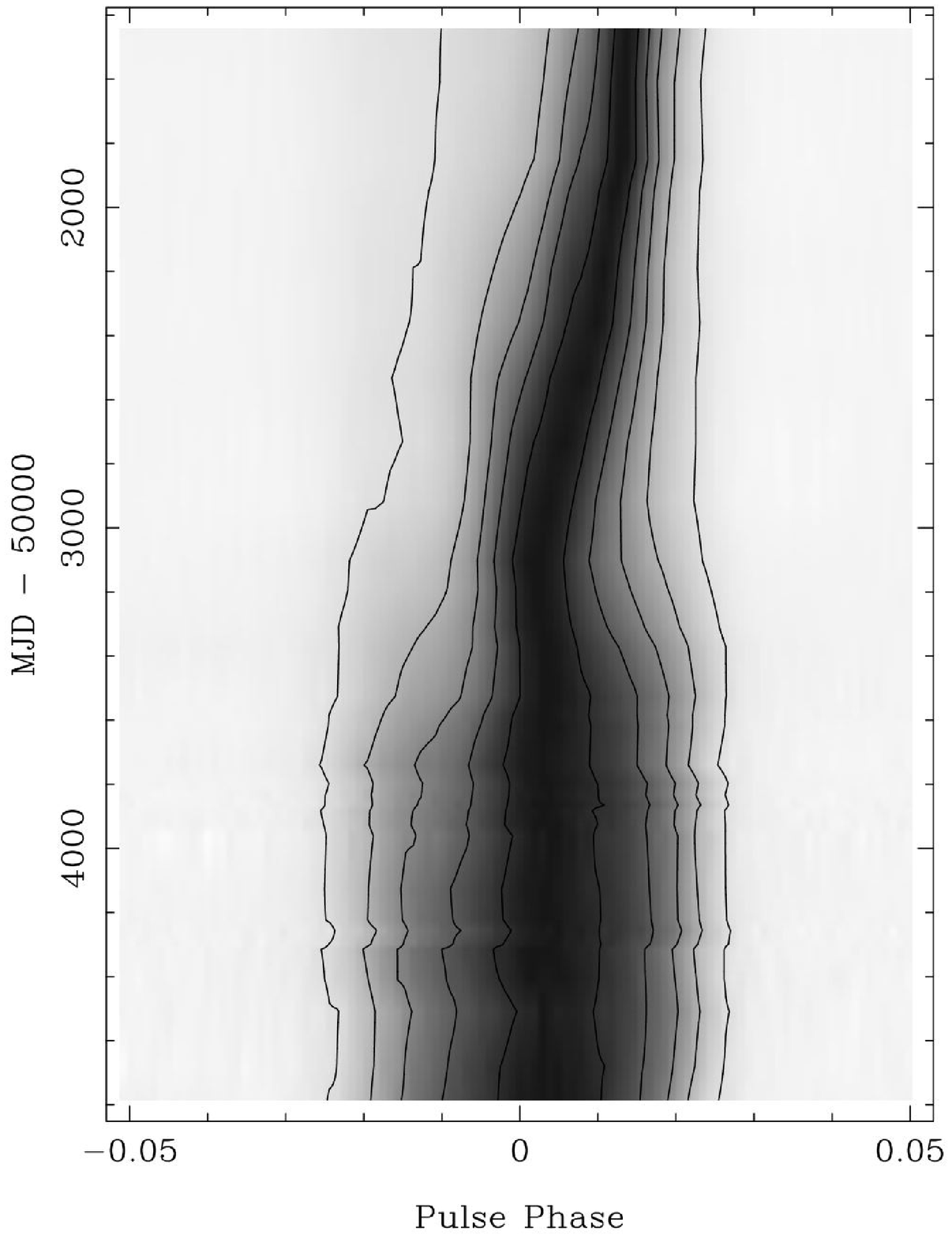}} &
\mbox{\includegraphics[width=75mm]{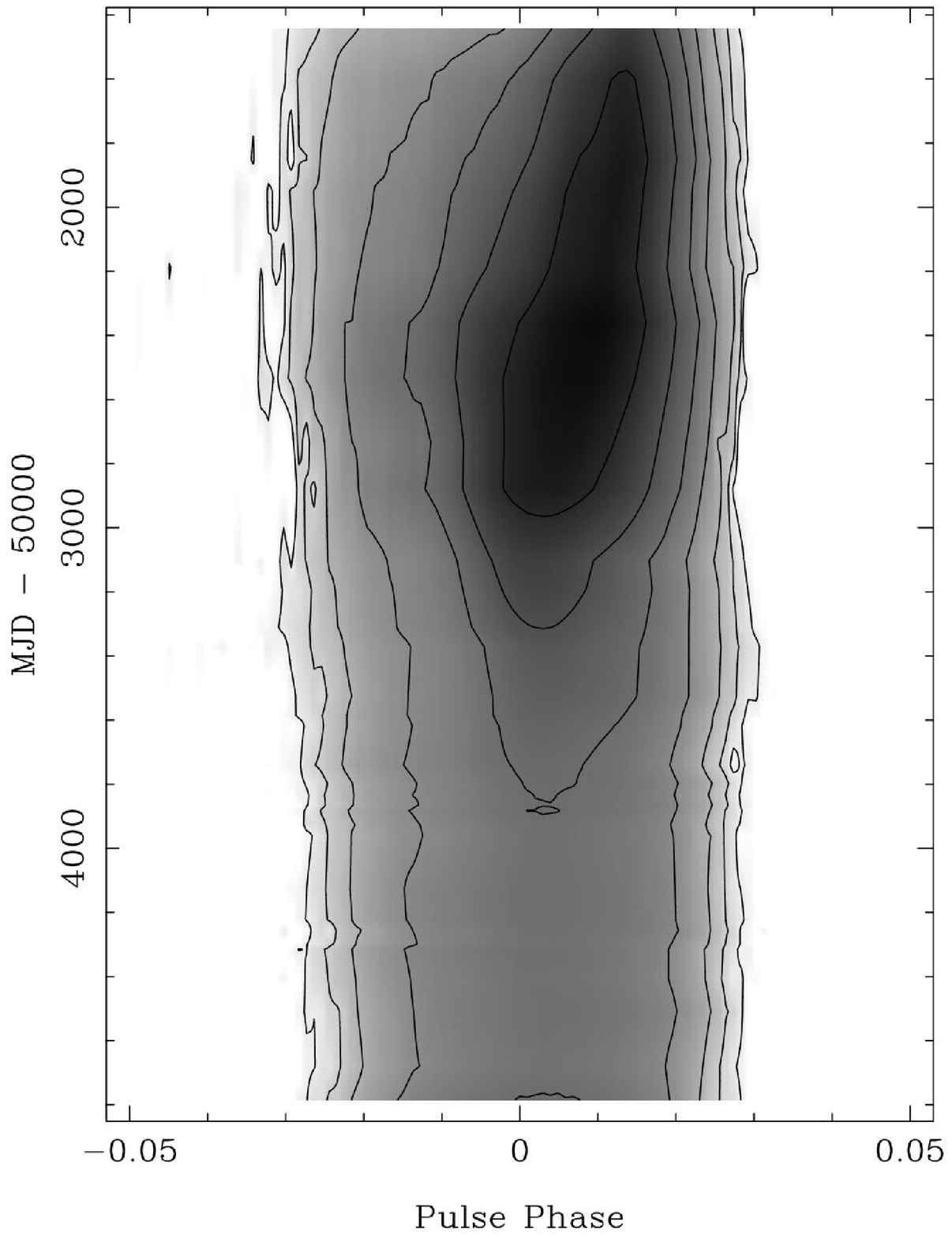}} 
\end{tabular}
\caption{Greyscale plots showing the time variation of the mean pulse
  profile for PSR J1141$-$6545. The greyscale linearly interpolates
  between the observed profiles. For the left-hand plot the profiles
  have been normalised to unity peak amplitude (as for
  Figure~\ref{fg:prf}) and the greyscale is linear with contour lines
  at 0.1, 0.3, 0.5, 0.7 and 0.9 of the profile peak. For the
  right-hand plot the profile amplitudes are in flux density units and
  the greyscale is logarithmic. The contour lines are at 0.01, 0.02,
  0.05, 0.1, 0.2 and 0.5 of the largest profile peak.}\label{fg:grey}
\end{center}
\end{figure*}

Average total-intensity profiles as a function of time are shown in
Figure~\ref{fg:prf} illustrating the profile evolution\footnote{Data
  from some closely spaced epochs have been averaged to improve
  clarity.} At early times the profile was dominated by a strong
trailing peak with a broad ramp of emission leading up to it, whereas
at late times, the profile is more symmetric with an approximately
Gaussian shape. Figure~\ref{fg:grey} shows the profile evolution in
the form of greyscale plots with both linear (left) and logarithmic
(right) intensity scales. The logarithmic plot clearly illustrates a
striking property of the profile evolution, that the width at a very
low level (the lowest contour is at 1\% of the peak) is remarkably
constant, suggesting that the low-level flux-density contours
represent the overall beam extent. As will be further described in
Section~\ref{sec:model}, the outer parts of the profile are relatively
stable, both in shape and flux density. The profiles in
Figures~\ref{fg:prf} and \ref{fg:grey} have therefore been aligned
using the midpoint at a constant flux density of 2\% of the maximum as
a reference point. This is an important assumption as it affects not
only the interpretation of the various profile components, but also
the parameters derived from any timing analysis. With this assumption,
the peak of the profile has moved in a somewhat step-wise fashion from
near its trailing edge to near its center. This evolution will be
interpreted in terms of the growth and decay of components
representing bright regions of the beam in Section~\ref{sec:model}
below.

Figure~\ref{fg:poln} shows the mean pulse polarization profiles for
PSR J1141$-$6545 at two epochs separated by about three years. Plotted
PAs refer to the band-center frequency as marked on the
plots.\footnote{Polarization plots for PSR J1141$-$6545 given by
  \citet{hbo05} have the opposite sign of $V$ and the opposite sense
  of position-angle swing to the IAU and pulsar conventions used for
  observational results in this paper. We also note that the same
  comments apply to the polarization profiles for the Double Pulsar,
  PSR J0737$-$3039A/B, presented by \citet{drb+04} and
  \citet{hbo05a}. As a consequence, the rotation measures given in
  these latter papers have the wrong sign. The current best estimate
  of the RM for PSR J0737$-$3039A/B with the corrected sign is
  $+112.3\pm 1.5$~rad~m$^{-2}$ \citep{drb+04}.}  Dramatic changes are
observed over the three-year span in the (total intensity) pulse
profile and in the polarization parameters. Once the sign differences
are allowed for and ignoring the absolute value of PA, the 2004
profile in Figure~\ref{fg:poln} agrees well with the profile given by
\citet{hbo05} taken about two months earlier. In particular, the PA
changes rapidly near the profile center, indicating that our line of
sight traverses the beam relatively close to the magnetic
axis. However there are clear and signficant departures from the PA
variation expected for the simple RVM.

\begin{figure*}
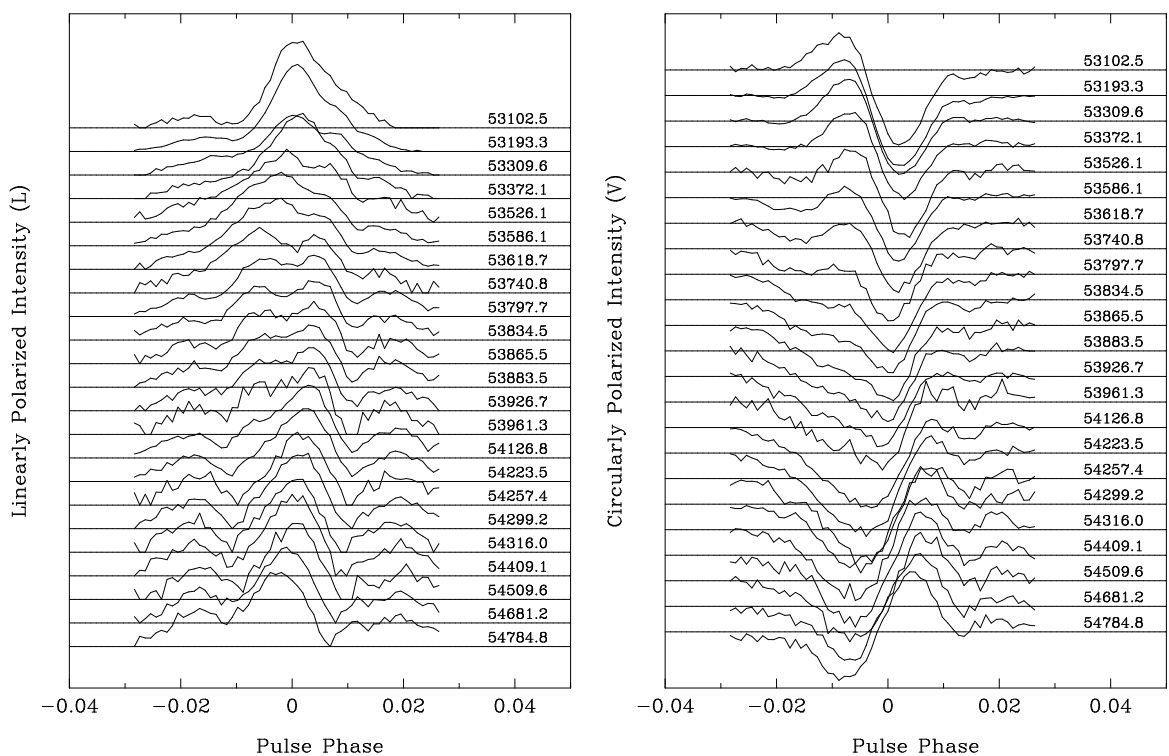

\begin{center} 
\begin{tabular}{cc}
\mbox{\includegraphics[width=75mm]{lline.ps}} &
\mbox{\includegraphics[width=75mm]{vline.ps}} 
\end{tabular}
\caption{Evolution of the linearly polarized intensity $L = (U^2 +
  Q^2)^{1/2}$ (left) and Stokes $V$ (right) profiles for PSR
  J1141$-$6545. The mean MJD for each profile is shown. }\label{fg:lvpol}
\end{center}
\end{figure*}

The evolution of the profiles of linearly polarized intensity $L$ and
of Stokes $V$ across the pulse over the 4.5-year span of the
polarization data are shown in Figure~\ref{fg:lvpol}. Both show
systematic changes as a function of time. In the case of the $L$
profiles, components at different pulse phases appear to get stronger
or weaker as a function of time, whereas the variation of Stokes $V$
is more consistent across the whole profile. It is striking that the
variation of $V$ across the pulse changes smoothly from a positive to
negative sign change at early epochs to a negative to positive change
at late epochs. Especially at later times, there are clear dips in the
linearly polarized intensity at pulse phases near $\pm 0.01$. As will
be discussed further below, these are attributed to overlapping of
approximately orthogonally polarized components.

Figure~\ref{fg:pa} shows the evolution of the PA variation across the
pulse. The observed PA variations are clearly not well described by
the RVM and there are clear systematic trends in the PA variations as
a function of time. While there are some differences between adjacent
PA profiles which may in part result from residual calibration errors
or low-level radio-frequency interference, in general the trends in
polarization properties with time are smooth and consistent despite
the use of different receivers and backend systems at different times,
again showing that our calibration procedures are effective. 

\begin{figure*}
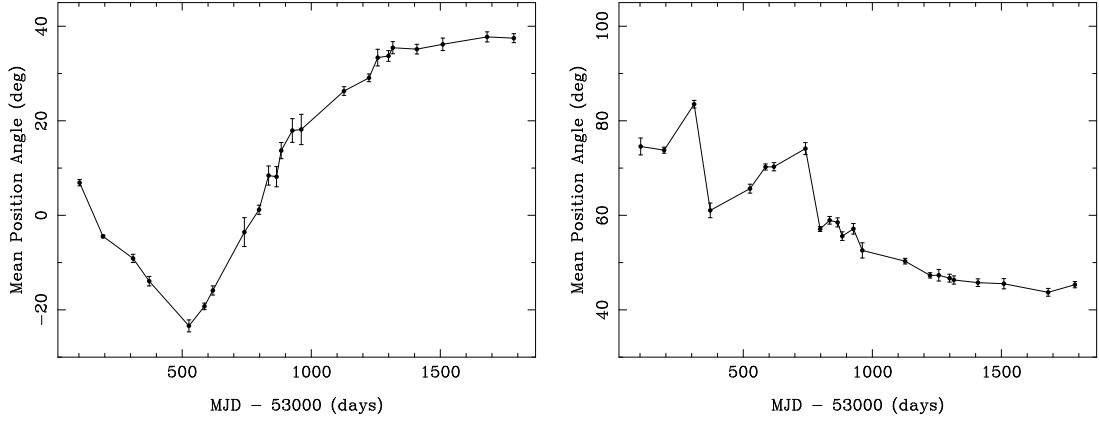

\begin{center} 
\begin{tabular}{cc}
\mbox{\includegraphics[width=55mm,angle=-90]{pa_centre.ps}} &
\mbox{\includegraphics[width=55mm,angle=-90]{pa_outer.ps}} 
\end{tabular}
\caption{Evolution of the mean absolute position angle (in the IAU
  convention) for the centre (left) and outer (right) parts of the PSR
  J1141$-$6545 pulse profile at the reference frequency of 1400
  MHz. The y axes in the two plots have the same
  scale. }\label{fg:pa_evol}
\end{center} 
\end{figure*}

Most striking is the clear difference in PA evolution between the
central and the outer parts of the profile. Figure~\ref{fg:pa_evol}
shows the PA variations over the 4.5-year data span averaged over the
inner part (phase $-0.0078$ to 0.0068) and the outer parts ($-0.0253$
to $-0.0117$ and 0.0136 to 0.0253) of the profile. Variations of the
mean PA for the inner part of the pulse profile are very significant
with an intial decrease followed by a rapid increase which slows at
later epochs. For the outer parts of the profile, the mean PA at early
times shows some fluctuations but overall is consistent with a slow
decrease, flattening at later times. Especially at early times, the
polarization in the outer parts of the profile is quite weak and hence
the fluctuations in PA can be attributed to remaining systematic errors
and/or contributions from the wings of the stronger emission in the
center of the profile.

\begin{figure*}
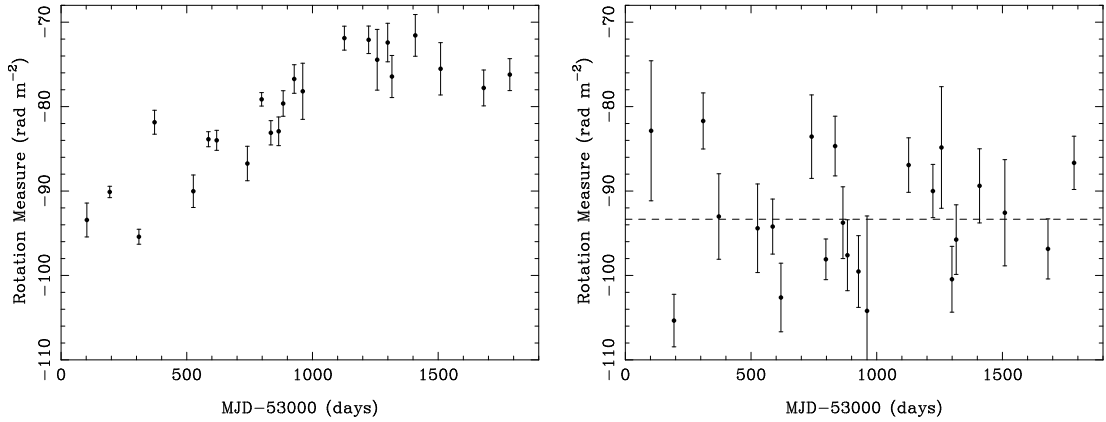

\begin{center} 
\begin{tabular}{cc}
\mbox{\includegraphics[width=55mm,angle=-90]{rm_centre.ps}} &
\mbox{\includegraphics[width=55mm,angle=-90]{rm_outer.ps}} 
\end{tabular}
\caption{Time variation of (apparent) rotation measures for the centre
  (left) and outer (right) parts of pulse profile. The dashed line in
  the righthand plot shows the weighted mean RM.  }\label{fg:rm}
\end{center} 
\end{figure*}

These PA changes could be due to variations in RM or to aspect changes
resulting from the precessional motion. In Figure~\ref{fg:rm} we show
the RMs separately for the central and outer parts of the profile
computed using the iterative method described in
Section~\ref{sec:obs}. For the outer parts of the profile there is no
significant time variation. The weighted mean RM value is $-93.4\pm
2.2$ rad m$^{-2}$ with a reduced $\chi^2$ of 3.1; the quoted
uncertainty is the weighted rms deviation multiplied by
$\sqrt(\chi^2_r)$.\footnote{This RM value is used to rotate observed
  PAs to the required reference frequency.} In contrast, for the
central part, there is a clear systematic variation of measured RM
with time. We believe that this is not a true variation in the
interstellar RM -- it would be much larger than RM variations in other
pulsars that are attributed to a changing path through the
interstellar medium, e.g., for the Vela pulsar
\citep{hhc85}. Furthermore, it is inconsistent with the changes in the
central PA shown in Figure~\ref{fg:pa_evol}. Rather, we believe the
observed changes in both PA and apparent RM for the central part of
the PSR J1141$-$6545 profile are due to variations in the relative
amplitudes of observed profile components with different PAs and
spectral indices resulting from the changing aspect of the line of
sight relative to the beam axis as the pulsar spin axis
precesses. This is somewhat analogous to the apparent RM variations
(as a function of pulse phase) seen in PSR B2016+28 by \citet{rbr+04}
which are attributed to over-lapping non-orthogonal components. 

We adopt PAs in the outer zones as representing the underlying
magnetic field structure. Consequently, these PA variations should be
consistent with the RVM. The two sides are consistently at
approximately the same PA, as expected in the RVM. Also, in some
cases, e.g., the 2007 August 4 profile shown in Figure~\ref{fg:poln},
there is evidence for the PA variations expected for the RVM at the
inner edges of the outer PAs.  Since there is no signficant RM
variation for these parts, we interpret the observed slow decrease in
outer-zone PAs shown in the right panel of Figure~\ref{fg:pa_evol} as
intrinsic and resulting from precession of the pulsar spin axis.

\section{Modelling of the Precessional Changes}\label{sec:model}
\begin{figure}
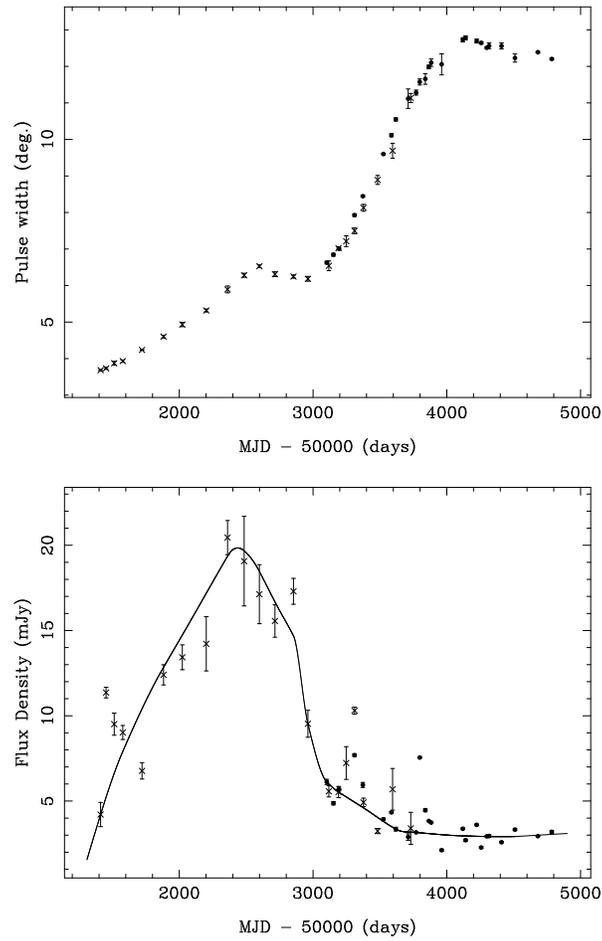

\begin{center} 
\begin{tabular}{c}
\mbox{\includegraphics[width=60mm,angle=-90]{w50_time.ps}} \\
\mbox{\includegraphics[width=60mm,angle=-90]{s14_time.ps}} 
\end{tabular}
\caption{Mean 50\% pulse widths and mean flux densities at 1400 MHz
  versus time for PSR J1141$-$6545. AFB data are marked with crosses
  and WBC and PDFB1 data are marked with dots. Error bars are
  $\pm 1$ standard deviation of the mean. The curve is a spline fit to
  average flux densities which represents the probable intrinsic
  variation.}\label{fg:w50s14}
\end{center}
\end{figure}

\begin{figure}
\includegraphics[width=150mm]{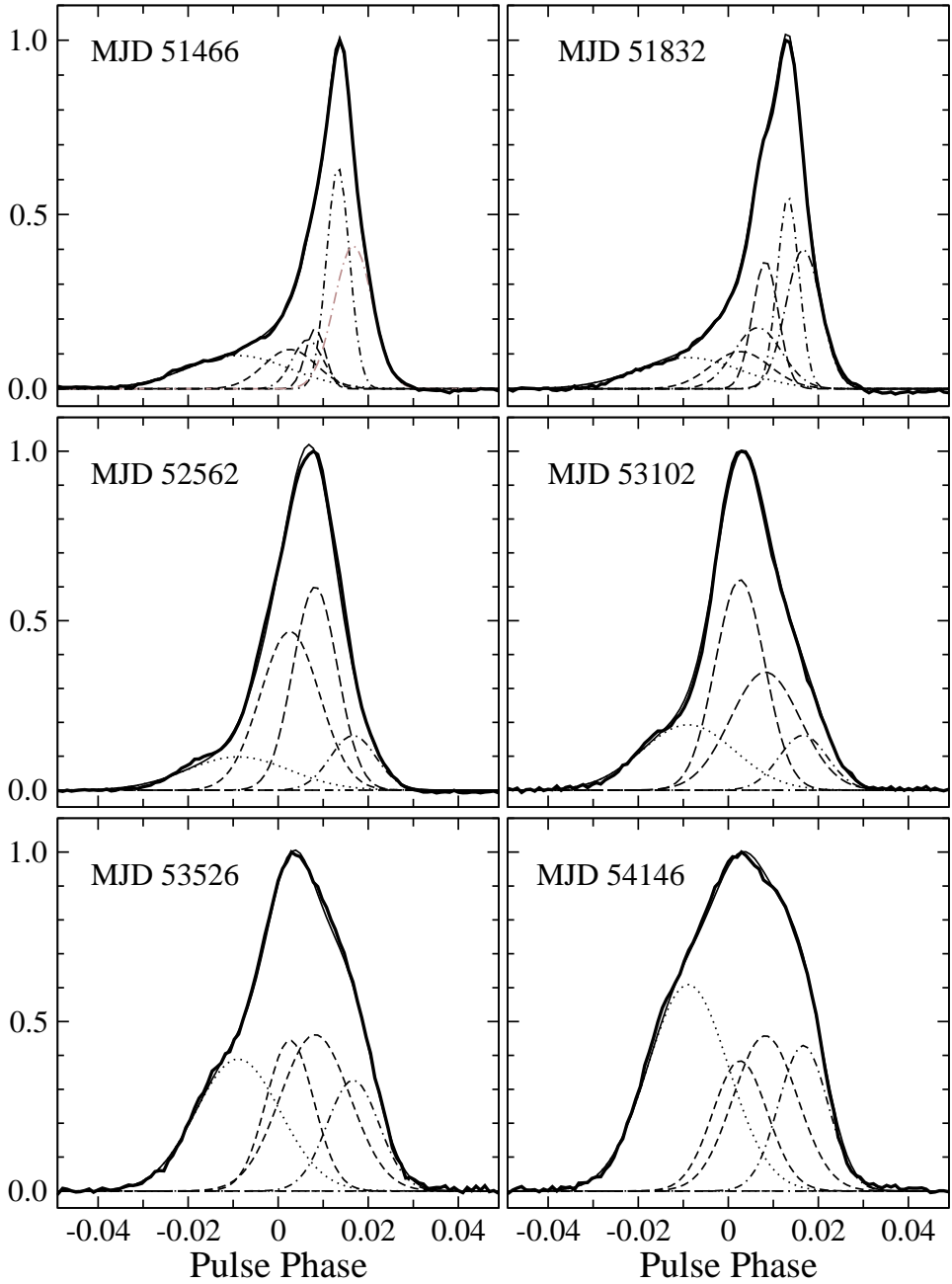}
\caption{A representative set of normalised total-intensity profiles
  for PSR J1141$-$6545, together with the fitted gaussian components
  at each epoch. The sum of the gaussian components is plotted as a
  thin line, but it is mostly hidden by the observed
  profiles.}\label{fg:prof}
\end{figure}

By modelling the observed PA variations in terms of precessional
motion of the pulsar's spin axis, we can infer additial properties of
the system, for example, the spin-orbit misalignment angle and the
precessional phase. As a first step in the modelling, we fit the
observed total-intensity profiles (Figure~\ref{fg:prf}) with gaussian
components. A total of six components are fitted, although for later
epochs only four are required. We have used fixed central phases for
the components; only the amplitude and width of each component is
allowed to vary as a function of epoch. The component phases, given in
Table~\ref{tb:comp_phs}, were chosen using an iterative process by
fitting the first and last observed profiles. Fitted components and
their sum are shown in Figure~\ref{fg:prof} for a selected set of
profiles which cover most of the observed data span, showing that an
excellent fit to the observed profiles is obtained. This retroactively
justifies our decision to fix the component
phases. Figure~\ref{fg:comps} gives the widths and mean flux densities
of the fitted components. The mean flux densities are scaled so that
their sum equals the smoothed flux-density variation shown in
Figure~\ref{fg:w50s14}. For components 3 and 5, the fitted amplitudes
dropped to zero after MJDs 52100 and 52500, respectively.

It is striking that the amplitude of each component varies smoothly
from one epoch to the next and that the widths of the components are
relatively stable. This suggests that these components represent real
physical zones of emission which are fixed in longitude on the
star. We note that most of the observed flux density variation comes
from the central and trailing components. Although the leading
component 1 is relatively weak at early epochs and dominates the
profile at late epochs, its absolute amplitude remains approximately
constant across the whole data span.

\begin{figure}
\includegraphics[width=120mm]{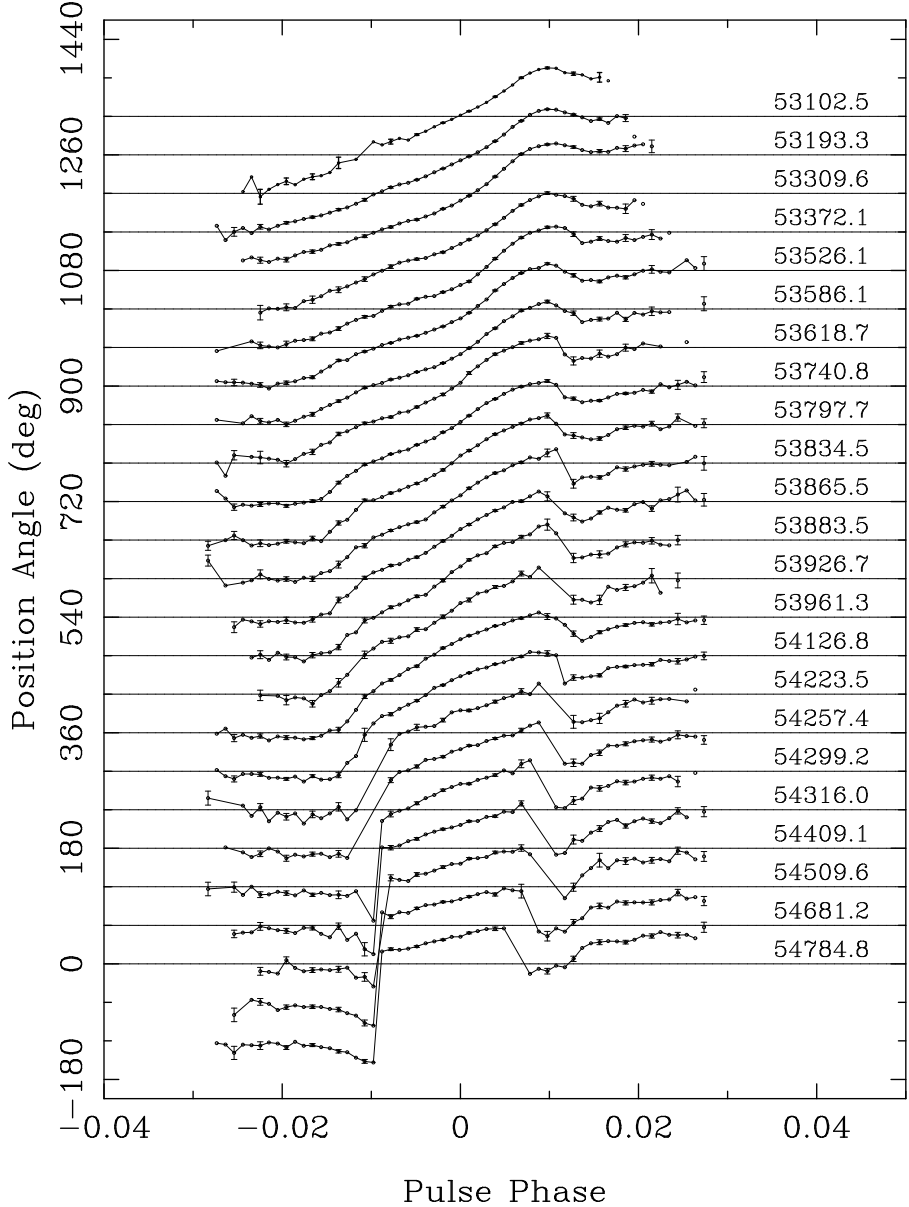}
\caption{Variations of linear position angle (PA) for PSR J1141$-$6545
  mean pulse profiles as a function of phase and time. Error bars
  ($\pm 1\sigma$) are plotted on every third point. Position angles
  are absolute according to the IAU convention and have been rotated
  to refer to a reference frequency of 1400 MHz assuming a rotation
  measure of $-93.4$~rad~m$^{-2}$ (see below). For the last few
  epochs, PAs in the leading outer zone have been offset downward by
  $180\degr$ to maintain continuity in the time-variation of PAs in
  this zone. The mean MJD for each profile is listed. }\label{fg:pa}
\end{figure}

As discussed above, the observed polarization PA and $L$ variations
(Figs~\ref{fg:pa} and \ref{fg:lvpol}) strongly suggest that these
components may be grouped into two overlapping and quasi-independent
emission zones, one dominant in the wings of the observed pulse and
the other dominant in the central regions. We note that similar
non-RVM variations (not accountable for by overlapping orthogonal
polarization modes) are seen in other short-period pulsars, for
example, PSR B1913+16 \citep{bcw91} and PSR B1534+12
\citep{aptw96,sta04}. The outer components (1 and 6) may be identified
with the outer zones in the PA and $L$ plots. As discussed above, the
PAs in these zones together are well fitted by a single RVM. Note
that, as discussed in Sections~\ref{sec:intro} and \ref{sec:obs}, for
modelling purposes, we have adopted a right-hand coordinate system in
which the sign of PA is reversed compared to the astronomical
convention used in Section~\ref{sec:results}. We also choose $\phi_0 =
0.0$ and hold this fixed for all epochs.

\begin{figure}
\includegraphics[width=120mm]{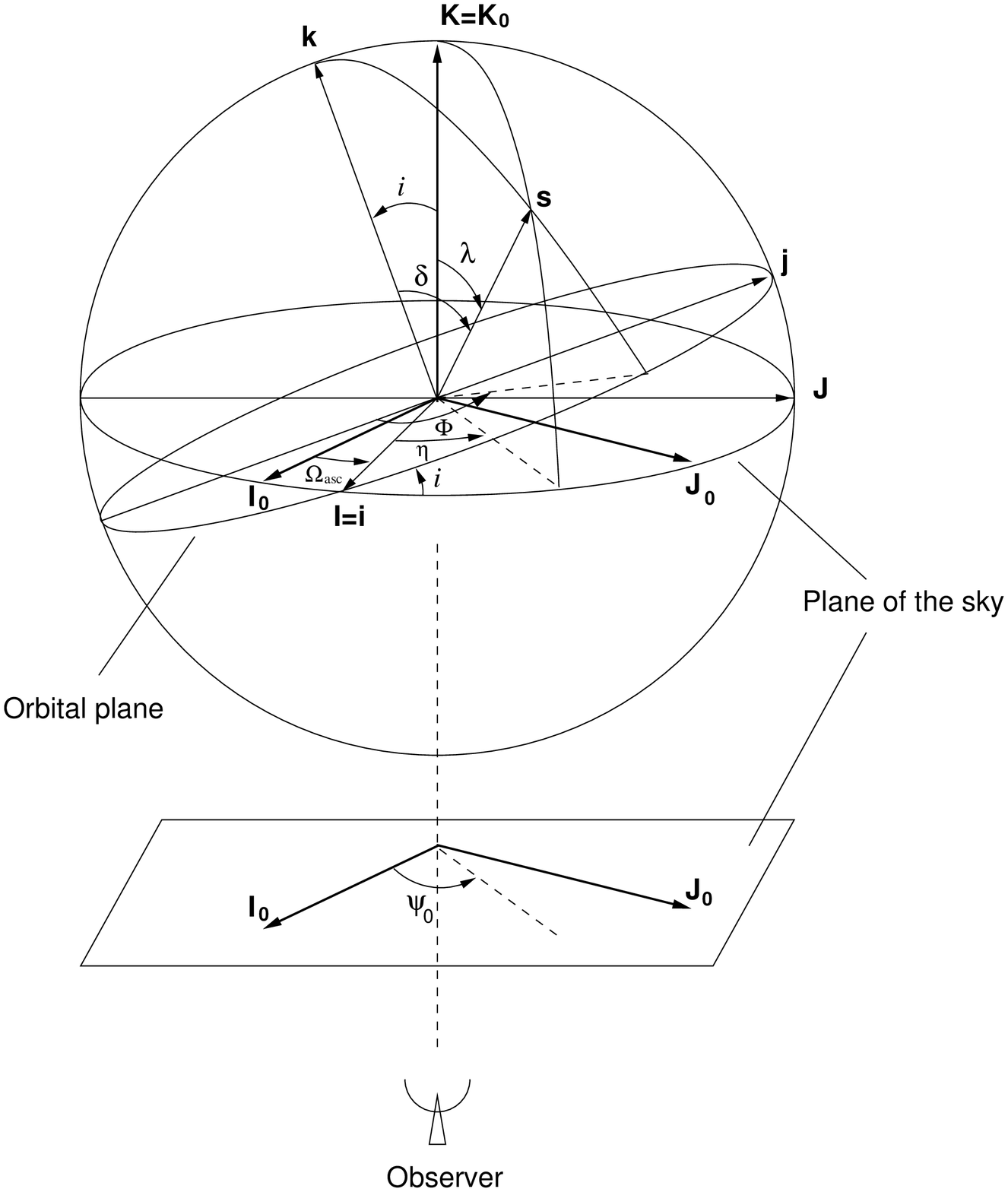}
\caption{Definition of angles used in the analysis of 
  precession of the pulsar spin axis (after DT92). The triad ($\bf
  I\;J\;K$) has $\bf K$ in the line-of-sight direction and the triad
  ($\bf i\;j\;k$) has $\bf k$ along the orbit normal, with $\bf I$ =
  $\bf i$ defined by the ascending node. $\bf I_0$ is toward East and
  $\bf J_0$ toward North on the sky, making a right-hand triad with
  $\bf K_0$ = $\bf K$. The pulsar spin vector $\bf s$ effectively
  precesses around the orbit normal $\bf k$ at the angular rate
  $\Omega_p$. }\label{fg:angles}
\end{figure}

Most previous discussions of relativistic spin precession in this and other pulsars
\citep[e.g.,][]{wrt89,kra98,sta04,hbo05} have concentrated on
measurement and interpretation of the rate of change of impact
parameter. For PSR J1141$-$6545, the impact parameter is not very well
determined because of the presence of the evidently independent
emission in the central part of the profile. However, since we have
measured absolute position angles, we have another, potentially more
sensitive, observable: the central PA of the RVM, $\psi_0$
\citep[cf.][]{sta04,kw09}. In the RVM, this is the direction of the spin
axis of the star projected on the sky plane (see
Figure~\ref{fg:angles}).  This of course varies as the pulsar precesses:
\begin{equation}\label{eq:psi0}
\psi_0 = \Omega_{asc} + \eta
\end{equation}
where $\Omega_{asc}$ is the longitude of the ascending node and
$\eta$, the precessional longitude, is the angle of the pulsar spin
axis relative to the ascending node projected on the sky plane. In
this equation, $\psi_0$ is the observed PA corrected to infinite frequency
using the measured RM of $-93.4$ rad m$^{-2}$. The polar angle $\lambda$
varies with the precession according to:
\begin{equation}\label{eq:lambda1}
\cos\lambda = \cos\delta \cos i - \sin\delta \sin i \cos\Phi
\end{equation}
\begin{equation}\label{eq:lambda2}
\cos\eta \sin\lambda = \sin\delta \sin\Phi
\end{equation}
where $\delta$ is the spin-orbit misalignment angle, $\Phi$ is the
precessional angle measured in the orbit plane from $\bf -j$ and
$\Phi_0$ is the value of $\Phi$ at $t=t_0$ (DT92) and
\begin{equation}\label{eq:Phi}
\Phi = \Omega_p (t-t_0) + \Phi_0.
\end{equation}
From Equation~(\ref{eq:psi0}), since $\Omega_{asc}$ is effectively
constant\footnote{The orbit does precess since it is the total angular
  momentum which is conserved. However, since the orbital angular
  momentum is much greater than the spin angular momentum of either
  star, the variation is very small and can be neglected in this
  analysis.}, the precessional variation in $\psi_0$ is determined by
the variation of $\eta$. From Equation~(\ref{eq:lambda1}), we can
determine $\cos\lambda$ as a function of time and the fixed angles
$\delta$, $i$ and $\Phi_0$.  We take $i = 73\fdg0$ from \citet{bbv08},
take $t = t_0$ at MJD 53000.0, near the center of the data span and
the beginning of our polarization monitoring, and take the GR
prediction for the spin precession rate $\Omega_p$, approximately
$1\fdg36$ yr$^{-1}$. Then, using Equation~\ref{eq:lambda2} (where
there is no sign ambiguity in $\sin\lambda$ since $0 \leq \lambda <
\pi$), we obtain $\cos\eta$. From the same spherical triangle used to
derive Equation~(\ref{eq:lambda2}), we also have
\begin{equation}\label{eq:cosdelta}
\cos\delta = \cos\lambda \cos i - \sin i \sin\lambda \sin\eta
\end{equation}
giving us $\sin\eta$ and hence $\eta$ as a function of time (and the
fixed but unknown angles $\Omega_{asc}$, $\delta$ and $\Phi_0$)
without further ambiguity \cite[cf.][]{kw09}. 

Consequently, the set of three parameters, $\delta$, $\Phi_0$,
$\Delta\psi_0$, (where $\Delta\psi_0$ absorbs the contribution to
$\psi_0$ by $\Omega_{asc}$ and any other intrinsic constant offset to
the PA values) describes the variation of $\psi_0$ as a function of
time. At the same time, the PA swing at each epoch is ideally
described by the RVM, which uses two further parameters: the magnetic
inclination $\alpha$ and the impact parameter $\beta$. (We have held
the central longitude of the RVM, $\phi_0$, at zero as discussed above.)
However, since
\begin{equation}\label{eq:lambda3}
\lambda = \pi - \zeta = \pi  - \alpha - \beta
\end{equation}
the angle $\beta$ can be computed from Eqn.~(\ref{eq:lambda1}) for a given
$\alpha$. As a result, only four parameters should describe the behavior of
the PA variations at all epochs.

As described above, the central range of the PA swing shows clear
deviations from an RVM. However, in a blind search for the best fit
solution, it nevertheless adds valuable information as it clearly
indicates (by its slope) that $\beta<0$ (in the DT92 convention). It
can also be noted that the slope steepens and then flattens again
towards the end of our dataset (see Figure ~\ref{fg:pa}). In order to
use the geometric information provided by the central part but to
still tie the results to the outer wings, we decided to decrease the
relative weight of the central part in the fitting process (by
multiplying its errors by a factor of five\footnote{Varying this
  factor between three and ten did not have a major impact on the
  results described below.}). We allowed a PA offset
between the central part relative to the RVM determined by the outer
wings, and this was fitted for independently for each epoch.  We take
longitudes in the range of $|\phi|\le3\degr$ (0.00833 in phase) to
define the central region, whereas PAs in the transition zone $3\degr
< |\phi| < 5\degr$ (0.00833 to 0.0139 in phase) were ignored in the
fit.

\begin{figure}
\includegraphics[width=90mm,angle=270]{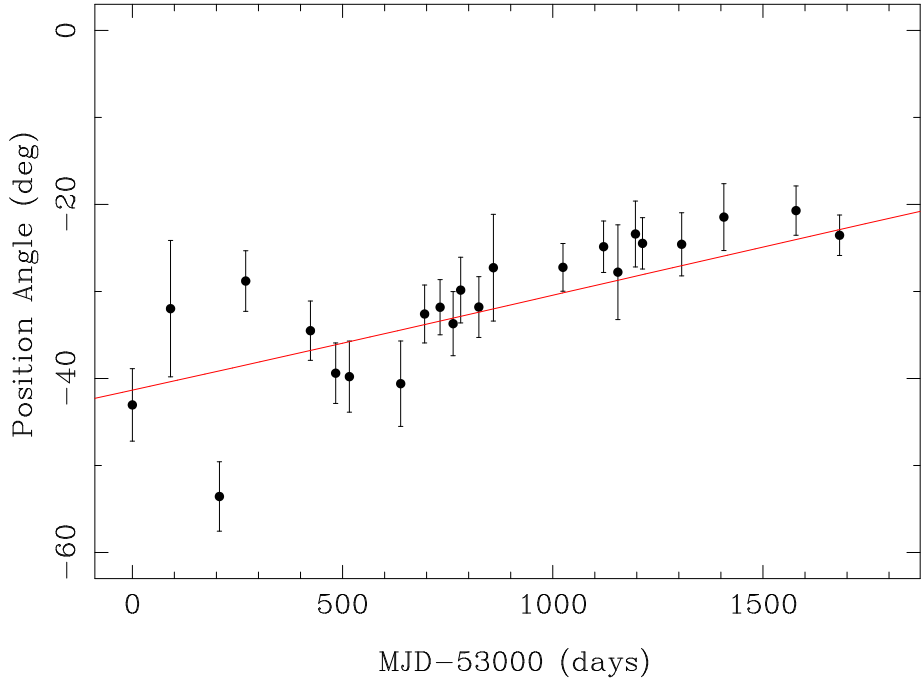}
\caption{Variations of the fitted central PA of the RVM fit ($\psi_0$)
  for the outer parts of the pulse profile. Note that the $\psi_0$ values are
  defined in the DT92 convention, are absolute and have been rotated to
  refer to infinite frequency assuming a rotation measure of
  $-93.4$~rad~m$^{-2}$. They therefore have the opposite sign to the astronomical
  PA plotted in Figure~\ref{fg:pa_evol}. The line is the expected
  variation in $\psi_0$ based on the best-fit precessional model.
}\label{fg:psi0}
\end{figure}

\begin{figure}
\includegraphics[width=120mm]{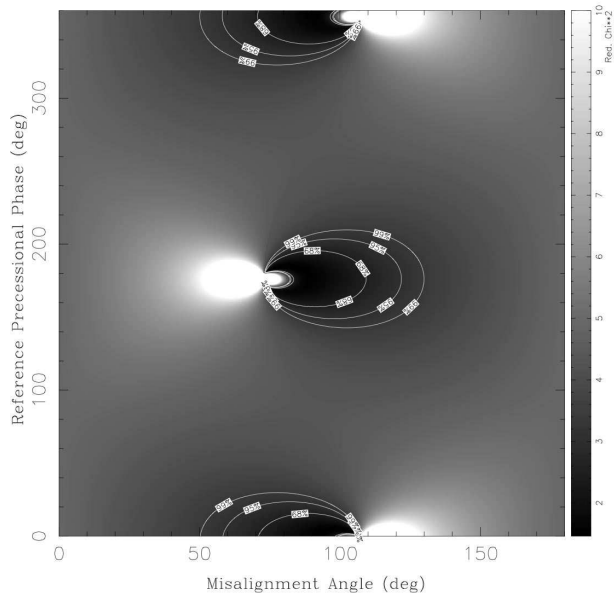}
\caption{Plot of $\chi^2$ in the spin-orbit misalignment angle
  ($\delta$) -- reference precessional phase ($\Phi_0$) plane for
  fits to the outer-zone RVM central PAs shown in
  Figure~\ref{fg:psi0}. Black and white correspond to a reduced
  $\chi^2$ values of 1.0 and 10.0 respectively. Contour lines at the
  68\%, 95\% and 99\% confidence levels around the two possible
  solutions are marked.}\label{fg:chi2pa0}
\end{figure}

As a first step of the fitting procedure, we constrain the allowable
values of $\delta$ and $\Phi_0$ using just the outer-zone PAs. For this
purpose, we compute the mean PA of the outer wings and fit its
behavior as a function of time (see Figure~\ref{fg:psi0}).  We use a
grid search in the $\delta$ -- $\Phi_0$ plane, where at each grid
point we compute the $\chi^2$ for the fit of the model values (based
on Equations~\ref{eq:psi0} -- \ref{eq:cosdelta}) to the computed
values of $\psi_0$ (Figure~\ref{fg:psi0}).
Figure~\ref{fg:chi2pa0} is a plot of $\chi^2$ in the $\delta$ --
$\Phi_0$ plane, showing two clear and well defined solutions. For this
plot, we have taken $i=73\fdg0$; the plot for $180-i$ is
mirror-symmetric.

The solution with $\Phi_0 \sim 0$ corresponds to $\lambda>\pi/2$ while
the other solution implies $\lambda < \pi/2$. We can use the
information provided by the PA swing to identify the appropriate
solution in the following way. From the PA swing (in the DT92
convention), the slope is negative and hence $\beta<0$. Also, the fact
that the PA becomes flat at the outer wings suggests an outer line of
sight (that is, the line of sight is on the equatorial side of the
magnetic axis when $\phi = \phi_0$, see, e.g., Lorimer \& Kramer
2005)\nocite{lk05} rather than an inner one. The combination of both
is only possible in a solution with $\zeta = \alpha+\beta > \pi/2$,
corresponding to solution where $\lambda = \pi - \zeta < \pi/2$. This
uniquely identifies the region around $\delta \sim 90\degr$ and
$\Phi_0 \sim 175\degr$ as the correct solution. 

\begin{figure}
\includegraphics[width=120mm]{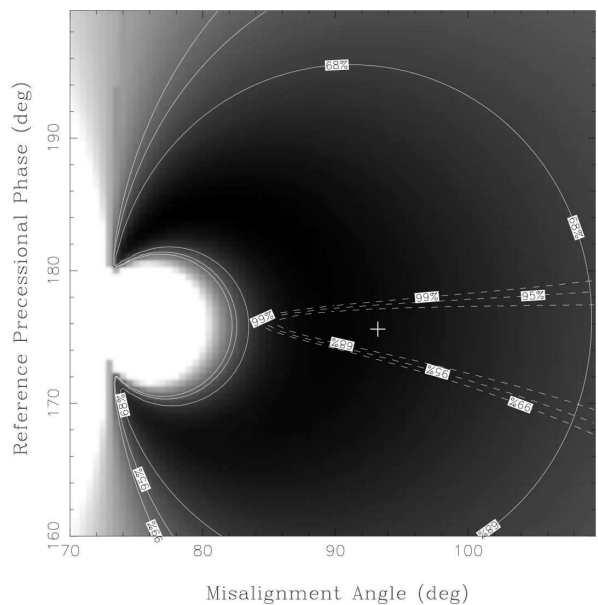}
\caption{Plot of $\chi^2$ in the spin-orbit misalignment angle
  ($\delta$) -- reference precessional phase ($\Phi_0$) plane around
  the region of the preferred solution. The greyscale and the solid
  line contours are identical to those in Figure~\ref{fg:chi2pa0}
  except for the change in scale.  The dashed contours are at the
  68\%, 95\% and 99\% confidence levels for the global RVM fit. The
  best-fit position shown by + corresponds to the variation in
  $\psi_0$ shown by the line in
  Figure~\ref{fg:psi0}.}\label{fg:chi2final}
\end{figure}

The best solution from this fitting procedure corresponds to $\lambda$
being very close to zero. We also know that $\beta$ is close to zero from
the rapid swing of PA in the central region and the flat PAs in the
outer zones which are only a few degrees from the pulse
center. Therefore, from Equation~(\ref{eq:lambda3}), we have that
$\alpha\sim\pi$. However, the total-intensity profiles and PA swing
show that the magnetic axis cannot be exactly aligned (or
counter-aligned) with the rotation axis. The fact that the sign of the
central PA swing remains the same at all epochs also shows that
$\beta$ does not change sign as a result of the precessional
motion. This suggests that the fit to the PA data as a whole will lie
somewhat to the right of the best-fit point shown in
Figure~\ref{fg:chi2pa0}.  This idea is confirmed by a blind fit to all
PA data using only the four parameters $\delta$, $\Phi_0$,
$\Delta\psi$ and $\alpha$. The search is again performed on a
$\delta$ -- $\Phi_0$ grid, where at each grid point we execute a simplex
algorithm to minimize for the other two parameters. The results
(collapsed in $\Delta\psi$ and $\alpha$ space) are shown in
Figure~\ref{fg:chi2final}. We show in grayscale the previous results
of the PA-offset fit and overlay contour levels for the global RVM fit
to the PA data. The model variation of $\psi_0$, shown in
Figure~\ref{fg:psi0}, has a reduced $\chi^2$ of 7.4  with 920 degrees of
freedom. Assumed and derived parameters for the system are given in
Table~\ref{tb:model}. Figure~\ref{fg:delta_pdf} shows the joint probability
density function for the spin-orbit misalignment angle $\delta$ derived from the
fits to the PA variations and the global RVM fit (Figure~\ref{fg:chi2final}).

Given the geometry of the system we can obtain the
model time variation of impact parameter $\beta$ using Equations
\ref{eq:lambda1} -- \ref{eq:Phi} and \ref{eq:lambda3}. The result,
given in Figure~\ref{fg:beta}, shows that the observer's line of sight
has approached the magnetic axis direction through most of our
observed data span, reaching its closest approach around MJD 54000,
and is now receding again. This reversal in the variation of $\beta$
is supported by the fact that the pulse width reached a maximum at
about MJD 54200 and is now decreasing (Figure~\ref{fg:w50s14}). Note
that we have obtained these results by fitting only the polarization
data, i.e., we did not make use of the total intensity
information. Although the over-all variation of pulse width is
consistent with the precessional model, the detailed variations are
more complex. This shows that the assumption of a uniform circular beam is
not appropriate.

\section{Shape of  the emission beam}\label{sec:beam}
\begin{figure}
\includegraphics[width=100mm,angle=270]{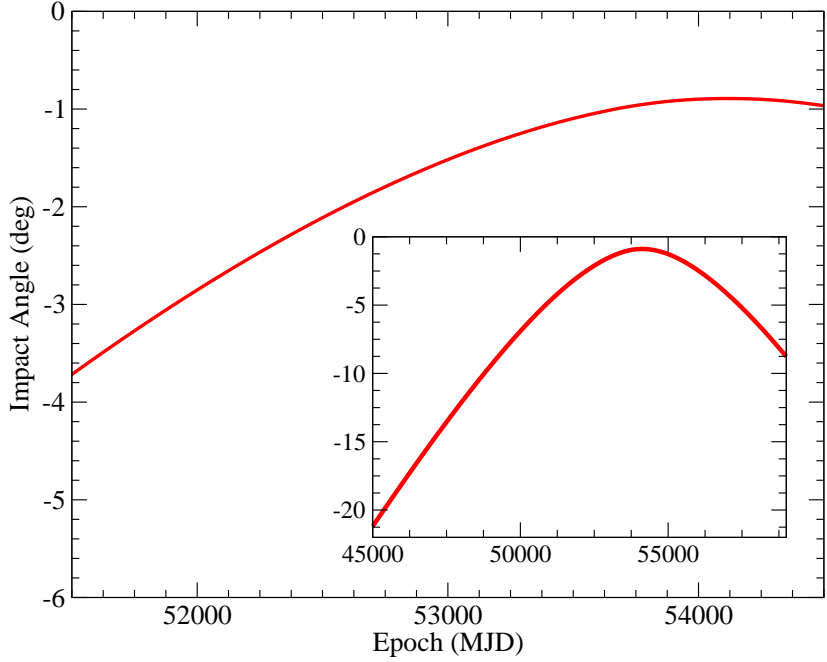}
\caption{Variation of impact parameter $\beta$ as a function time over
  the observed data span based on the final model parameters
  (Table~\ref{tb:model}). The inset shows an extended time
  range.}\label{fg:beta}
\end{figure}

\begin{figure}
\includegraphics[width=90mm,angle=-90]{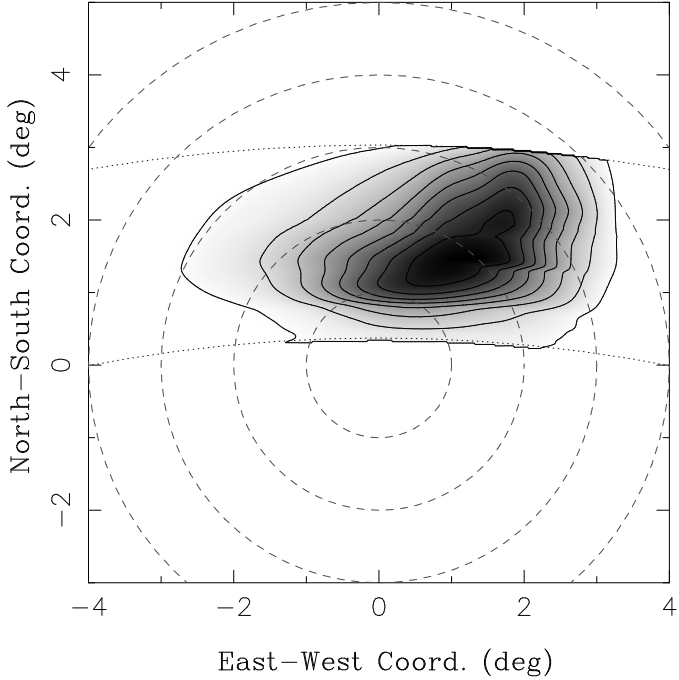}
\caption{Two-dimensional shape of the emission beam over the traversed
  region in a planar projection centered on the magnetic pole. The
  inclination of the magnetic axis ($\alpha$) is taken to be
  $160\degr$, the most probable value from our precessional
  solution. The coordinate system is centered on the magnetic pole and
  the dashed circles are at $1\degr$ intervals in radius. The dotted
  lines show the path traversed by the line of sight with the upper
  one corresponding to the beginning of the data span and the lower
  one to the minimum impact parameter reached around MJD 54000
  (Figure~\ref{fg:beta}). Since $\beta <0$ and $\zeta = \alpha +
  \beta$, the line-of-sight trajectory is above the magnetic
  pole. Furthermore, since $\alpha > 90\degr$, the line-of-sight
  trajectory is ``outer'', that is, on the equatorial side of the
  magnetic pole. The grey scale and contour lines are on a logarithmic
  scale with a factor of approximately 0.07 between successive
  contours. }\label{fg:beam}
\end{figure}

Given the variation of $\beta$ over the data span
(Figure~\ref{fg:beta}) and the time variation of component shapes and
amplitudes (Figure~\ref{fg:comps}), we can compute the two-dimensional intensity
profile of the emission beam over the traversed region. For each epoch
we compute the traverse of our line of sight across the polar region
and accumulate the pulse intensity in a two dimensional grid centered
on the magnetic axis. The resulting beam pattern is shown in
Figure~\ref{fg:beam}. Because of the reversal in the time-derivative of
$\beta$ when it was close to zero, we only see one half of the
polar-cap region.  Despite this we can clearly see that the beam is
quite asymmetric with no evidence for a core-cone or ring structure
that is symmetric about the magnetic pole. The partially filled beam
can be described as ``patchy'', albeit with just one major patch in
the region scanned so far. Although this is the first two-dimensional
map of an emission beam to clearly show such patchy structure, there
is good evidence that the emission beam from most pulsars is best
described in this way \citep{lm88,hm01}.

Although the observed pulse width is about average (the median pulse
width for all pulsars is about $10\degr$), the small value of
$180\degr - \alpha \sim 20\degr$ implies that the intrinsic beamwidth
is small. Figure~\ref{fg:beam} shows that the emitting region fits within a circle
of radius about $4\degr$ centered on the magnetic axis. Excluding millisecond pulsars,
observed pulse widths are generally consistent with the relation
\begin{equation} 
\rho = 6\fdg5 P^{-0.5}
\end{equation}
\citep{gks93,kwj+94}. For PSR J1141$-$6545 the predicted value of
$\rho$ is about $10\fdg3$, much larger than the emitting zone
traversed so far.  

PSR J1141$-$6545 was not detected in the Parkes 70cm survey
\citep{mld+96} although, even at the present relatively low
flux-density levels, a detection with signal/noise ratio of the order
of 50 would have been expected. Observations within half a beamwidth
of the pulsar position were made 1993 July 14 (MJD
49182). Figure~\ref{fg:beta} shows that the impact parameter $\beta$
at that time was about $-8\degr$. This non-detection therefore
suggests that the beam half-width in latitude is $\lesssim 8\degr$,
although this must be qualified because of the patchy beam
structure. There are good arguments \citep[e.g.,][]{nv83,man96} that
beams in young pulsars are elongated in the latitude direction and
hence more fan-like than circular. An elongated beam has also been
suggested for PSR B1913+16 \citep{wt02,cw08}. Observations over the
next decade or two will establish whether or not this is the case for
PSR J1141$-$6545 as $\beta$ returns to large (negative) values.

\section{Implications for the progenitor system}\label{sec:kick}

There is good and increasing evidence \citep[e.g.][]{jhv+05,nr08} that
the velocity and spin vectors of young pulsars are nearly aligned.
This suggests that the kick from the supernova (SN) explosion, the
dominant factor in determining the pulsar space motion, may also
determine the spin direction \citep[see also][]{sp98}.  Since
PSR~J1141$-$6545 is a young pulsar, its spin direction is therefore
likely to be aligned with the SN kick. While precession will have
caused the azimuthal angle of the pulsar spin relative to the
(post-SN) orbital angular momentum  to wrap many times over the
lifetime of the pulsar, the polar angle ($\delta$ in this paper)
should still reflect the direction between the kick and the post-SN
orbit normal. 

We have explored if this expectation can set any interesting
constraints on the kick imparted to the forming neutron star, using
the binary orbit and kick description in \citet{kal96} and
\citet{wkk00}.  In terms of their variables, we find:
\begin{equation}
  \cos\delta = \frac{ V_{\rm kz} (2 V_{\rm ky} +V_0)}{|V_{\rm k}| 
[V_{\rm kz}^2+(V_{\rm ky}+V_0)^2]^{1/2}}.
\end{equation}
We follow the procedure set out in \citet{tds05} and the Bayesian
version in \citet{std+06}, which follows the principles of
\citet{wkk00} but which puts priors on the system velocity, the angle
$\theta$ between the pre- and post-SN orbit normals\footnote{Note that
  this is the angle labeled $\delta$ in \citet{tds05} and
  \citet{std+06} and corresponds to the spin-orbit misalignment angle
  for a recycled pulsar whose companion has undergone a SN explosion.}
and the orientation $\Omega$ of the binary system on the sky.  Each
sampling of the priors determines the coefficients of a quadratic
equation in $m_{2i}$, the mass of the pre-SN star \citep[Equation 8
of][]{tds05}, which may or may not have a solution that falls within
acceptable ranges for the pre-SN mass and the size of the (assumed
circular) pre-SN orbit.  As this is a young system, we did not evolve
the system's motion back in time through the Galaxy to identify
plausible birth sites \citep{wkk00}, nor did we evolve the orbital
size and eccentricity to account for gravitational radiation losses
\citep{pet64}, but simply assumed the current velocity and orbital
parameters reflect the birth properties. Note that the results are
therefore independent of the assumed distance; we assumed a distance
of 3.7~kpc \citep{obv02}.  We put uniform priors on $\theta$ (sampling
from $0\degr$ to $180\degr$ and testing both positive and negative
values in each trial) and $\Omega$ ($0\degr$ to $360\degr$). We
sampled the companion mass over ($1.02 \pm 0.01$)~M$_{\odot}$ and the
total system mass over ($2.28911\pm 0.00026$)~M$_{\odot}$, based on
the orbital parameters reported in \citet{bbv08}. We subtracted these
to get the pulsar mass for the given trial, and used the mass function
computed from the orbit presented in \citet{bbv08},
0.176550265~M$_{\odot}$, to derive $\sin i$. We considered only the
cases where $\cos i > 0$ in order to match the modeling in the
previous section.

\begin{figure}
\includegraphics[width=120mm,angle=270]{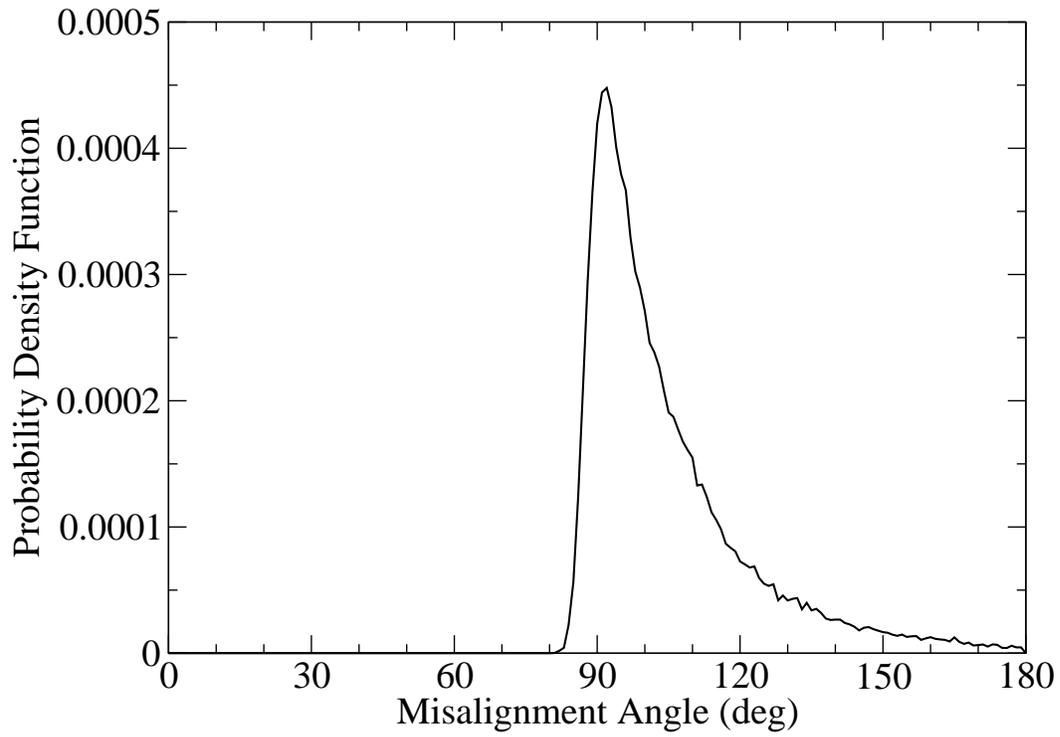}
\caption{Final probability distribution for the spin-orbit misalignment angle
  $\delta$ derived from the central PA and global RVM fits. }
  \label{fg:delta_pdf}
\end{figure}

\begin{figure*}
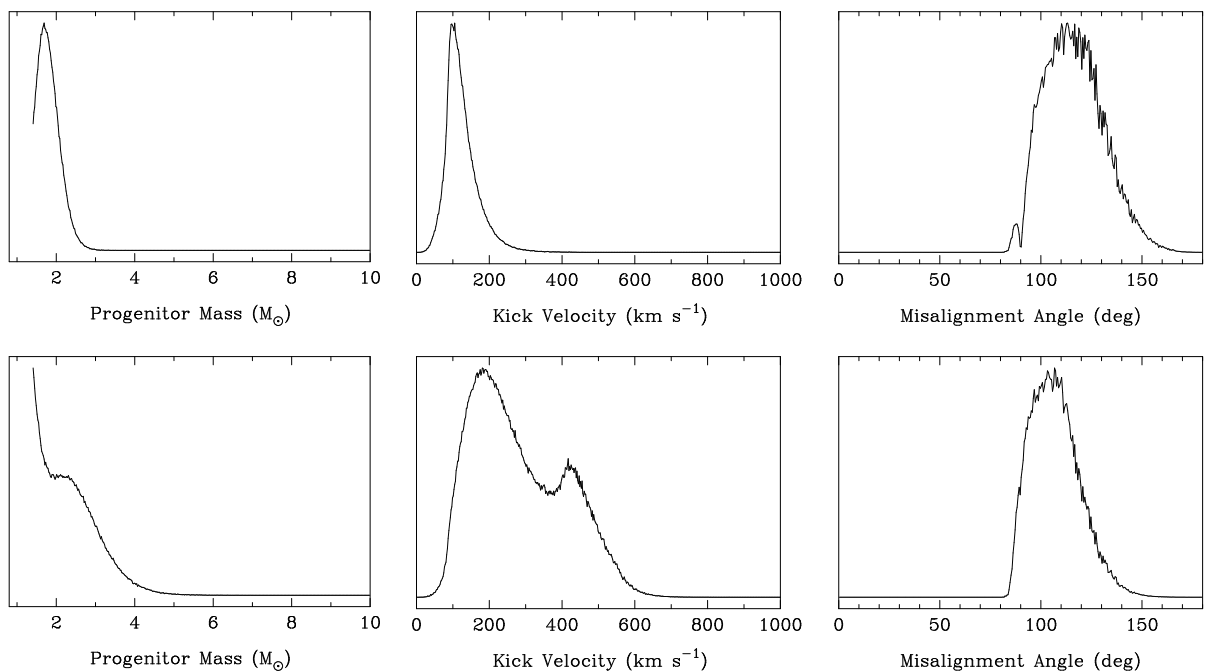

\begin{center} 
\begin{tabular}{ccc}
\mbox{\includegraphics[width=42mm,angle=-90]{pdf_mass_50c.ps}} &
\mbox{\includegraphics[width=42mm,angle=-90]{pdf_kick_50c.ps}} &
\mbox{\includegraphics[width=42mm,angle=-90]{pdf_delta_50c.ps}} \\
\mbox{\includegraphics[width=42mm,angle=-90]{pdf_mass_100c.ps}} &
\mbox{\includegraphics[width=42mm,angle=-90]{pdf_kick_100c.ps}} &
\mbox{\includegraphics[width=42mm,angle=-90]{pdf_delta_100c.ps}} 
\end{tabular}
\caption{Relative probability density functions for the mass of the
  exploding star immediately before the supernova, the magnitude of
  the resulting kick velocity and the post-SN spin-orbit misalignment angle of
  the pulsar spin ($\delta$) after taking into account the constraints
  on $\delta$ from the PA model fitting. The upper row
  corresponds to a one-dimensional dispersion of 50 km s$^{-1}$ in the
  assumed distribution of post-SN system velocity and the lower row to
  a dispersion of 100 km s$^{-1}$. }\label{fg:pdfc}
\end{center} 
\end{figure*}

While there is a scintillation velocity measurement for this system
\citep{obv02} which could in principle provide some constraints on the
angle between the proper motion and $\Omega$, similar measurements
have been shown to be quite unreliable in the Double Pulsar, probably
because of the effects of anisotropies in the interstellar medium
\citep{rkr+04,cmr+05,ksm+06} and hence we do not make use of this
information here.  Instead we investigate Maxwellian velocity
distributions with 1-dimensional dispersions of 50\,km~s$^{-1}$ and
100\,km~s$^{-1}$ (these dispersions are of order the transverse
velocity derived from the scintillation measurements), combined with a second
Maxwellian with 1-dimensional dispersion of 12\,km~s$^{-1}$
representing the pre-SN peculiar velocity.  We then derive Bayesian
probability distributions for the pre-SN and kick parameters as
described in \citet{std+06}, considering pre-SN masses in the range
1.4 -- 10.0~M$_{\odot}$ to allow for the binding energy of the neutron
star \citep{lp01}.  For each acceptable solution of the quadratic
equation, the relevant set of parameters is assigned a uniform
likelihood \citep[as in][]{std+06} for a ``natural'' weighting. We
then make use of the probability distribution for the post-SN pulsar
spin-orbit misalignment angle $\delta$ obtained from the PA fitting
(Figure~\ref{fg:delta_pdf}) to give a constrained weighting on the
range of acceptable pre-SN parameters and kick velocity.  The derived
probability density functions for the pre-SN stellar mass, the kick
velocity and the spin-orbit misalignment angle after constraining with the
allowed misalignment-angle distribution (Figure~\ref{fg:delta_pdf})
are shown in Figure~\ref{fg:pdfc}. Table~\ref{tb:pdf} gives the median
values of the parameters and the limits at 68\% and 95\%
confidence. We note that the natural weighting tends to disfavour
spin-orbit misalignment angles around $90\degr$, so the constrained distribution
is biased toward slightly larger angles compared to Figure~\ref{fg:delta_pdf}.

Not unexpectedly, the assumed distribution of post-SN system
velocities has a significant effect on the likely kick velocity, with
larger values requiring a larger kick. However, the most probable kick
velocity is relatively small in both cases, about 100 km s$^{-1}$ for a
dispersion of 50 km s$^{-1}$ and about 180 km s$^{-1}$ for a
dispersion of 100 km s$^{-1}$. Median velocities (Table~\ref{tb:pdf})
are somewhat higher since the distributions have a high-velocity
tail. Never-the-less, they are still small compared with mean or
median of the inferred three-dimensional pulsar velocity distribution
\citep{ll94,hllk05,fk06}.

Relatively low progenitor masses are strongly favoured with
median values of about 2 M$_\odot$. The pulsar mass implied by the
timing fits for companion mass and total mass \citep{bbv08} is just
1.27~M$_\odot$, lower than average \citep{tc99}, but very similar to
the mass of PSR J0737$-$3039B, the second-born and slow pulsar in the
Double Pulsar system. This system also has a relatively low implied
kick velocity from the B pulsar formation
\citep[e.g.][]{ps05,wkf+06,std+06} but see also \citet{kvw08}. These low
masses and kick velocities may imply formation of the neutron star in
an electron-capture collapse of the core of an ONeMg white dwarf
\citep{pdl+05} but other evolutionary histories are also possible
\citep[e.g.,][]{dp03b}.

The inferred spin-orbit misalignment angle $\delta$ in the PSR J1141$-$6545
system is large, with a most probable value of between $100\degr$ and
$110\degr$. That $\delta$ is not small \citep[in contrast to pulsar A of the
Double Pulsar system,][]{std+06,fsk+08}, is consistent with the major
changes that we see in the pulse profile as a result of the precession
and in fact results from the fitting of the precessional model to
these changes.

\section{Summary and Conclusions}\label{sec:summary}
The dramatic long-term changes in pulse shape and amplitude observed
for PSR J1141$-$6545 (Figures~\ref{fg:w50s14} and \ref{fg:prf}) are
unprecedented in pulsar astronomy. Pulse profile changes are observed
in mode-changing pulsars \citep[e.g.][]{wmj07} and in radio emission
from magnetars \citep[e.g.][]{ccr+07} but these are of a quite
different character to the variations seen in PSR J1141$-$6545 and are
certainly caused by fluctuations in the emitted pulsar beam. Slow and
systematic variations similar to those observed in PSR J1141$-$6545
are seen in the pulse profiles of PSR B1913+16 and PSR B1534+12, but
these are of much smaller amplitude. All three of these pulsars are in
close binary orbits with massive companions. Consequently, the
observed profile variations most probably result from precession of
the pulsar spin axis resulting from spin-orbit coupling changing our
view of the pulsar beam.

\begin{figure}
\begin{tabular}{c}
\mbox{\includegraphics[width=100mm,angle=-90]{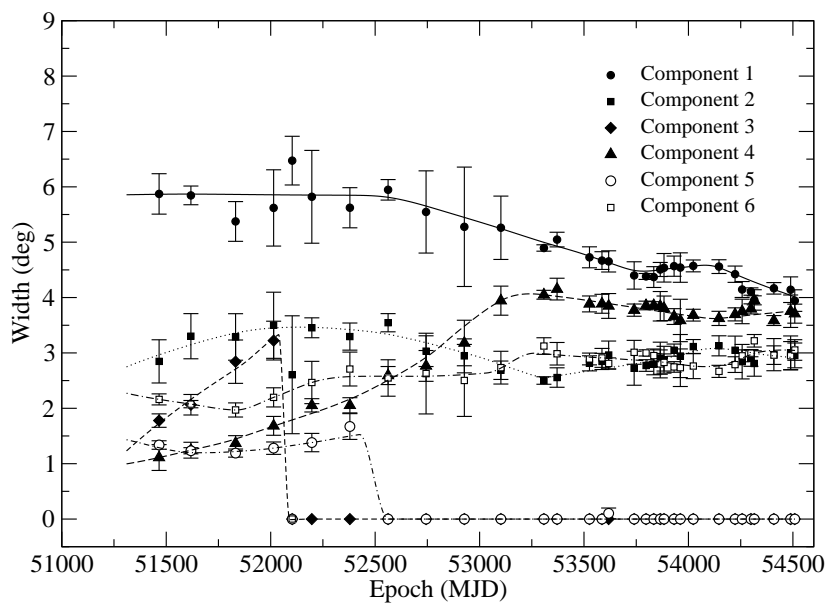}} \\
\mbox{\includegraphics[width=100mm,angle=-90]{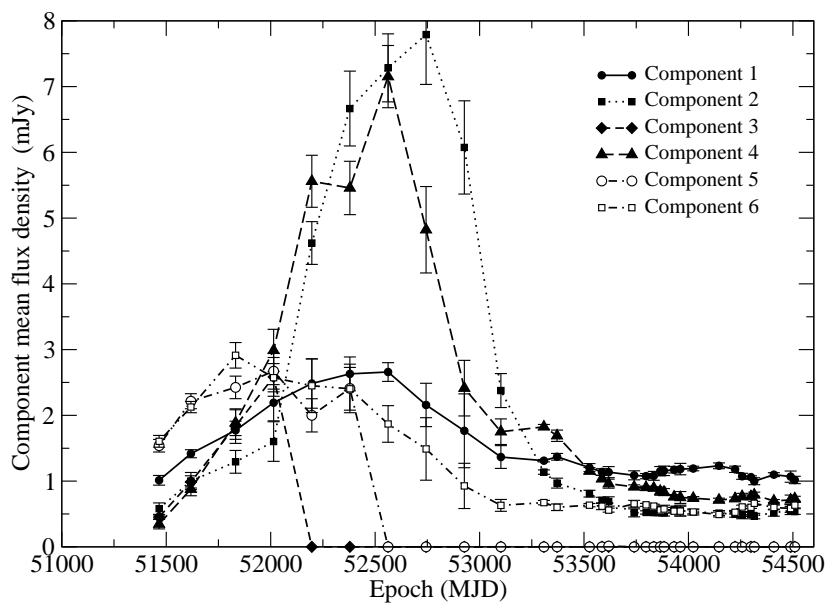}}
\end{tabular}
\caption{Time variations of the fitted widths and amplitudes of the
  six gaussian components. The amplitudes have been converted to
  component mean flux densities using $S = \sqrt{\pi}*a*w/P$, where
  $a$ is the relative component amplitude and $w$ is its width, and then scaled
  so that the mean flux density summed over all components equals the
  smoothed value in Figure~\ref{fg:w50s14} at each epoch. }\label{fg:comps}
\end{figure}

Precession of the pulsar's spin axis has resulted in large and
systematic changes in both the amplitude and the shape of the observed
pulse profile as our line of sight scanned across the polar
region. There are clearly large variations in emissivity across the
nominal beam area, consistent with the idea of a partially filled or
patchy beam. The large variations in the observed mean pulse profile
raise the issue of profile alignment when comparing different
epochs. Figure~\ref{fg:grey} shows that the pulse width is relatively
stable at the lowest contour levels. We have chosen to align the
profiles according to the mid-point of a constant flux-density level
near their extreme wings. This choice is motivated by the fact that
the outer components are relatively stable, both in flux density and
shape (Figure~\ref{fg:comps}). With this choice, alignment of the
polarization features over the data span is also very stable, giving
added support to our method of defining the absolute pulse phase
relative to the star. We note that this choice also defines a
reference phase for pulse timing. It is interesting to note that, for
PSR B1913+16 also, the pulse width defined by the lower beam contours
is more constant than, say, the pulse width at the 50\% level
\citep{wt05}.

Major changes are also observed in the pulse polarization parameters
over the 4.5 years in which the polarization has been monitored. There
has been a steady evolution of the circular polarization with the sign
of the sense reversal near the profile center changing from
positive-to-negative to negative-to-positive over the data
span. Variations in the linear polarization are complex with quite
different behavior in the central and outer zones of the pulse profile
(Figure~\ref{fg:pa}). Our choice of taking the outer-zone PAs to
represent the underlying magnetic-field structure is supported by a
number of factors. First, the outer-zone PAs are well fitted by the
rotating vector model (RVM) whereas this is less true of the
central-zone PAs. More importantly, there is a large and variable
offset between the central PAs and the outer-zone PAs. The time
variation of the mean central-zone PA is rapid and inconsistent with
precessional motion. Furthermore there is an apparent time variation
of RM for the central zone which is not present for the outer
zones. We attribute the complex behavior of the central-zone
polarization to the overlapping of independent components which have
different PAs and different spectral indices. It may be possible to
model the polarization variations based on the variations of the
total-intensity components shown in Figure~\ref{fg:comps} but this is
beyond the scope of the present work. We note that similar but much
less dramatic non-RVM variations are observed in the central parts of
the profiles for PSR B1913+16 and PSR B1534+12.

Fitting of a precessional model to the observed PA variations leads to
the conclusion that the spin-orbit misalignment angle $\delta$ is very large in
this system, with the pulsar spin axis nearly orthogonal to the
orbital angular momentum vector. This is consistent with the large
profile shape and polarization changes as the spin axis
precesses. Unfortunately, because of the complex and poorly understood
emission physics, it is not possible to turn the argument around and
use these observations as a test of the precessional predictions of
general relativity. It is interesting to note though that a large
spin-orbit misalignment angle is quite possible despite the low progenitor mass
and modest kick velocity. This result is not in conflict with the
conclusion of \citet{hbo05}, that the spin-orbit misalignment angle is probably
less than $30\degr$, since that was on the proviso that the pulsar was
not at a precessional phase when the angle between the observer's line
of sight and the spin axis ($\lambda$ in our terminology) was changing
rapidly. In our solution, $\lambda$ is close to zero and is changing
rapidly.

Our observational data span encompassed the time when the precessional
longitude passed through $180\degr$. This means that the impact
parameter $\beta$ reached an extremum during our data span. In fact,
$|\beta|$ reached a minimum value very close to zero at about MJD
54000 (Figure~\ref{fg:beta}). This solution is consistent with the
observed maximum in the pulse width at around the same time and implies that
we will only ever traverse one side of the emission beam. The slope of
the $\beta$ time-variation has now reversed and so we are now
retracing our earlier path across the beam. We therefore predict that
over the coming decade we will see a time-reversed ``replay'' of the
recent profile amplitude and shape evolution.

\section*{Acknowledgments} 
We thank our colleagues of the Parkes Multibeam Pulsar Survey team for
assistance with the observations reported in this paper. We also thank
Steve Thorsett and Rachel Dewey for their contributions to the
modelling of the effects of the pulsar kick velocity
(Section~\ref{sec:kick}). Pulsar research at UBC is supported by an
NSERC Discovery Grant and a CFI New Opportunities Grant to IHS and M.\
Berciu. CAM held an NSERC USRA and IHS an NSERC UFA for part of this
work. IHS further acknowledges support from the ATNF Distinguished
Visitors Program and the Swinburne University Visiting Distinguished
Researcher Scheme. The Parkes radio telescope is part of the Australia
Telescope which is funded by the Commonwealth Government for operation
as a National Facility managed by CSIRO.


\begin{thebibliography}{78}
\expandafter\ifx\csname natexlab\endcsname\relax\def\natexlab#1{#1}\fi

\bibitem[{Arzoumanian(1995)}]{arz95}
Arzoumanian, Z. 1995, PhD thesis, Princeton University

\bibitem[{Arzoumanian {et~al.}(1996)Arzoumanian, Phillips, Taylor, \&
  Wolszczan}]{aptw96}
Arzoumanian, Z., Phillips, J.~A., Taylor, J.~H., \& Wolszczan, A. 1996, ApJ,
  470, 1111

\bibitem[{{Bailes} {et~al.}(2003){Bailes}, {Ord}, {Knight}, \&
  {Hotan}}]{bokh03}
{Bailes}, M., {Ord}, S.~M., {Knight}, H.~S., \& {Hotan}, A.~W. 2003, ApJ, 595,
  L49

\bibitem[{Barker \& O'Connell(1975)}]{bo75}
Barker, B.~M. \& O'Connell, R.~F. 1975, ApJ, 199, L25

\bibitem[{{Bassa} {et~al.}(2008){Bassa}, {Wang}, {Cumming}, \&
  {Kaspi}}]{bwck08}
{Bassa}, C., {Wang}, Z., {Cumming}, A., \& {Kaspi}, V.~M., eds. 2008, {40 Years
  of Pulsars: Millisecond Pulsars, Magnetars and More}, Vol. 983 (New York:
  American Institute of Physics)

\bibitem[{{Bhat} {et~al.}(2008){Bhat}, {Bailes}, \& {Verbiest}}]{bbv08}
{Bhat}, N.~D.~R., {Bailes}, M., \& {Verbiest}, J.~P.~W. 2008, Phys. Rev. D, 77,
  124017

\bibitem[{Blaskiewicz {et~al.}(1991)Blaskiewicz, Cordes, \& Wasserman}]{bcw91}
Blaskiewicz, M., Cordes, J.~M., \& Wasserman, I. 1991, ApJ, 370, 643

\bibitem[{{Breton} {et~al.}(2008){Breton}, {Kaspi}, {Kramer}, {McLaughlin},
  {Lyutikov}, {Ransom}, {Stairs}, {Ferdman}, {Camilo}, \& {Possenti}}]{bkk+08}
{Breton}, R.~P., {Kaspi}, V.~M., {Kramer}, M., {McLaughlin}, M.~A., {Lyutikov},
  M., {Ransom}, S.~M., {Stairs}, I.~H., {Ferdman}, R.~D., {Camilo}, F., \&
  {Possenti}, A. 2008, Science, 321, 104

\bibitem[{{Burgay} {et~al.}(2005){Burgay}, {Possenti}, {Manchester}, {Kramer},
  {McLaughlin}, {Lorimer}, {Stairs}, {Joshi}, {Lyne}, {Camilo}, {D'Amico},
  {Freire}, {Sarkissian}, {Hotan}, \& {Hobbs}}]{bpm+05}
{Burgay}, M., {Possenti}, A., {Manchester}, R.~N., {Kramer}, M., {McLaughlin},
  M.~A., {Lorimer}, D.~R., {Stairs}, I.~H., {Joshi}, B.~C., {Lyne}, A.~G.,
  {Camilo}, F., {D'Amico}, N., {Freire}, P.~C.~C., {Sarkissian}, J.~M.,
  {Hotan}, A.~W., \& {Hobbs}, G.~B. 2005, ApJ, 624, L113

\bibitem[{{Camilo} {et~al.}(2007){Camilo}, {Cognard}, {Ransom}, {Halpern},
  {Reynolds}, {Zimmerman}, {Gotthelf}, {Helfand}, {Demorest}, {Theureau}, \&
  {Backer}}]{ccr+07}
{Camilo}, F., {Cognard}, I., {Ransom}, S.~M., {Halpern}, J.~P., {Reynolds}, J.,
  {Zimmerman}, N., {Gotthelf}, E.~V., {Helfand}, D.~J., {Demorest}, P.,
  {Theureau}, G., \& {Backer}, D.~C. 2007, ApJ, 663, 497

\bibitem[{{Clifton} \& {Weisberg}(2008)}]{cw08}
{Clifton}, T. \& {Weisberg}, J.~M. 2008, ApJ, 679, 687

\bibitem[{Coles {et~al.}(2005)Coles, McLaughlin, Rickett, Lyne, \&
  Bhat}]{cmr+05}
Coles, W.~A., McLaughlin, M.~A., Rickett, B.~J., Lyne, A.~G., \& Bhat, N. D.~R.
  2005, ApJ, 623, 392

\bibitem[{Damour \& Deruelle(1985)}]{dd85}
Damour, T. \& Deruelle, N. 1985, Ann. Inst. H. Poincar\'e (Physique
  Th\'eorique), 43, 107

\bibitem[{Damour \& Deruelle(1986)}]{dd86}
---. 1986, Ann. Inst. H. Poincar\'e (Physique Th\'eorique), 44, 263

\bibitem[{Damour \& Ruffini(1974)}]{dr74}
Damour, T. \& Ruffini, R. 1974, Academie des Sciences Paris Comptes Rendus
  Ser.\,Scie.\,Math., 279, 971

\bibitem[{Damour \& Taylor(1992)}]{dt92}
Damour, T. \& Taylor, J.~H. 1992, Phys. Rev. D, 45, 1840

\bibitem[{{Demorest} {et~al.}(2004){Demorest}, {Ramachandran}, {Backer},
  {Ransom}, {Kaspi}, {Arons}, \& {Spitkovsky}}]{drb+04}
{Demorest}, P., {Ramachandran}, R., {Backer}, D.~C., {Ransom}, S.~M., {Kaspi},
  V., {Arons}, J., \& {Spitkovsky}, A. 2004, ApJ, 615, L137

\bibitem[{Dewey \& Cordes(1987)}]{dc87}
Dewey, R.~J. \& Cordes, J.~M. 1987, ApJ, 321, 780

\bibitem[{{Dewi} \& {Pols}(2003)}]{dp03b}
{Dewi}, J.~D.~M. \& {Pols}, O.~R. 2003, MNRAS, 344, 629

\bibitem[{{Everett} \& {Weisberg}(2001)}]{ew01}
{Everett}, J.~E. \& {Weisberg}, J.~M. 2001, ApJ, 553, 341

\bibitem[{{Faucher-Gigu{\`e}re} \& {Kaspi}(2006)}]{fk06}
{Faucher-Gigu{\`e}re}, C.-A. \& {Kaspi}, V.~M. 2006, ApJ, 643, 332

\bibitem[{{Ferdman} {et~al.}(2008){Ferdman}, {Stairs}, {Kramer}, {Manchester},
  {Lyne}, {Breton}, {McLaughlin}, {Possenti}, \& {Burgay}}]{fsk+08}
{Ferdman}, R.~D., {Stairs}, I.~H., {Kramer}, M., {Manchester}, R.~N., {Lyne},
  A.~G., {Breton}, R.~P., {McLaughlin}, M.~A., {Possenti}, A., \& {Burgay}, M.
  2008, in  \cite{bwck08}, 474--478, 474

\bibitem[{Gil {et~al.}(1993)Gil, Kijak, \& Seiradakis}]{gks93}
Gil, J.~A., Kijak, J., \& Seiradakis, J.~H. 1993, A\&A, 272, 268

\bibitem[{Hamilton {et~al.}(1985)Hamilton, Hall, \& Costa}]{hhc85}
Hamilton, P.~A., Hall, P.~J., \& Costa, M.~E. 1985, MNRAS, 214, 5{P}

\bibitem[{{Han} \& {Manchester}(2001)}]{hm01}
{Han}, J.~L. \& {Manchester}, R.~N. 2001, MNRAS, 320, L35

\bibitem[{Han {et~al.}(2006)Han, Manchester, Lyne, Qiao, \& van
  Straten}]{hml+06}
Han, J.~L., Manchester, R.~N., Lyne, A.~G., Qiao, G.~J., \& van Straten, W.
  2006, ApJ, 642, 868

\bibitem[{Hobbs {et~al.}(2005)Hobbs, Lorimer, Lyne, \& Kramer}]{hllk05}
Hobbs, G., Lorimer, D.~R., Lyne, A.~G., \& Kramer, M. 2005, MNRAS, 360, 974

\bibitem[{{Hobbs} {et~al.}(2006){Hobbs}, {Edwards}, \& {Manchester}}]{hem06}
{Hobbs}, G.~B., {Edwards}, R.~T., \& {Manchester}, R.~N. 2006, MNRAS, 369, 655

\bibitem[{{Hotan} {et~al.}(2005{\natexlab{a}}){Hotan}, {Bailes}, \&
  {Ord}}]{hbo05}
{Hotan}, A.~W., {Bailes}, M., \& {Ord}, S.~M. 2005{\natexlab{a}}, ApJ, 624, 906

\bibitem[{{Hotan} {et~al.}(2005{\natexlab{b}}){Hotan}, {Bailes}, \&
  {Ord}}]{hbo05a}
---. 2005{\natexlab{b}}, MNRAS, 362, 1267

\bibitem[{{Hotan} {et~al.}(2004){Hotan}, {van Straten}, \&
  {Manchester}}]{hvm04}
{Hotan}, A.~W., {van Straten}, W., \& {Manchester}, R.~N. 2004, PASA, 21, 302

\bibitem[{{Johnston} {et~al.}(2005){Johnston}, {Hobbs}, {Vigeland}, {Kramer},
  {Weisberg}, \& {Lyne}}]{jhv+05}
{Johnston}, S., {Hobbs}, G., {Vigeland}, S., {Kramer}, M., {Weisberg}, J.~M.,
  \& {Lyne}, A.~G. 2005, MNRAS, 364, 1397

\bibitem[{Kalogera(1996)}]{kal96}
Kalogera, V. 1996, ApJ, 471, 352

\bibitem[{{Kalogera} {et~al.}(2008){Kalogera}, {Valsecchi}, \&
  {Willems}}]{kvw08}
{Kalogera}, V., {Valsecchi}, F., \& {Willems}, B. 2008, in {40 Years of
  Pulsars: Millisecond Pulsars, Magnetars and More}, ed. C.~{Bassa}, Z.~{Wang},
  A.~{Cumming}, \& V.~M. {Kaspi}, Vol. 983 (New York: American Institute of
  Physics), 433--441

\bibitem[{Kaspi {et~al.}(2000)Kaspi, Lyne, Manchester, Crawford, Camilo, Bell,
  D'Amico, Stairs, McKay, Morris, \& Possenti}]{klm+00a}
Kaspi, V.~M., Lyne, A.~G., Manchester, R.~N., Crawford, F., Camilo, F., Bell,
  J.~F., D'Amico, N., Stairs, I.~H., McKay, N. P.~F., Morris, D.~J., \&
  Possenti, A. 2000, ApJ, 543, 321

\bibitem[{{Kaspi} {et~al.}(2004){Kaspi}, {Ransom}, {Backer}, {Ramachandran},
  {Demorest}, {Arons}, \& {Spitkovskty}}]{krb+04}
{Kaspi}, V.~M., {Ransom}, S.~M., {Backer}, D.~C., {Ramachandran}, R.,
  {Demorest}, P., {Arons}, J., \& {Spitkovskty}, A. 2004, ApJ, 613, L137

\bibitem[{{Kramer}(1998)}]{kra98}
{Kramer}, M. 1998, ApJ, 509, 856

\bibitem[{{Kramer} {et~al.}(2006){Kramer}, {Stairs}, {Manchester},
  {McLaughlin}, {Lyne}, {Ferdman}, {Burgay}, {Lorimer}, {Possenti}, {D'Amico},
  {Sarkissian}, {Hobbs}, {Reynolds}, {Freire}, \& {Camilo}}]{ksm+06}
{Kramer}, M., {Stairs}, I.~H., {Manchester}, R.~N., {McLaughlin}, M.~A.,
  {Lyne}, A.~G., {Ferdman}, R.~D., {Burgay}, M., {Lorimer}, D.~R., {Possenti},
  A., {D'Amico}, N., {Sarkissian}, J.~M., {Hobbs}, G.~B., {Reynolds}, J.~E.,
  {Freire}, P.~C.~C., \& {Camilo}, F. 2006, Science, 314, 97

\bibitem[{{Kramer} \& {Wex}(2009)}]{kw09}
{Kramer}, M. \& {Wex}, N. 2009, Classical and Quantum Gravity, 26, 073001

\bibitem[{Kramer {et~al.}(1994)Kramer, Wielebinski, Jessner, Gil, \&
  Seiradakis}]{kwj+94}
Kramer, M., Wielebinski, R., Jessner, A., Gil, J.~A., \& Seiradakis, J.~H.
  1994, A\&AS, 107, 515

\bibitem[{{Lattimer} \& {Prakash}(2001)}]{lp01}
{Lattimer}, J.~M. \& {Prakash}, M. 2001, ApJ, 550, 426

\bibitem[{Lorimer \& Kramer(2005)}]{lk05}
Lorimer, D.~R. \& Kramer, M. 2005, Handbook of Pulsar Astronomy (Cambridge
  University Press)

\bibitem[{Lyne {et~al.}(2004)Lyne, Burgay, Kramer, Possenti, Manchester,
  Camilo, McLaughlin, Lorimer, D'Amico, Joshi, Reynolds, \& Freire}]{lbk+04}
Lyne, A.~G., Burgay, M., Kramer, M., Possenti, A., Manchester, R.~N., Camilo,
  F., McLaughlin, M.~A., Lorimer, D.~R., D'Amico, N., Joshi, B.~C., Reynolds,
  J., \& Freire, P. C.~C. 2004, Science, 303, 1153

\bibitem[{Lyne \& Lorimer(1994)}]{ll94}
Lyne, A.~G. \& Lorimer, D.~R. 1994, Nature, 369, 127

\bibitem[{Lyne \& Manchester(1988)}]{lm88}
Lyne, A.~G. \& Manchester, R.~N. 1988, MNRAS, 234, 477

\bibitem[{Manchester(1996)}]{man96}
Manchester, R.~N. 1996, in Pulsars: Problems and Progress, {IAU} Colloquium
  160, ed. S.~Johnston, M.~A. Walker, \& M.~Bailes (San Francisco: Astronomical
  Society of the Pacific), 193--196

\bibitem[{{Manchester} {et~al.}(2005){Manchester}, {Kramer}, {Possenti},
  {Lyne}, {Burgay}, {Stairs}, {Hotan}, {McLaughlin}, {Lorimer}, {Hobbs},
  {Sarkissian}, {D'Amico}, {Camilo}, {Joshi}, \& {Freire}}]{mkp+05}
{Manchester}, R.~N., {Kramer}, M., {Possenti}, A., {Lyne}, A.~G., {Burgay}, M.,
  {Stairs}, I.~H., {Hotan}, A.~W., {McLaughlin}, M.~A., {Lorimer}, D.~R.,
  {Hobbs}, G.~B., {Sarkissian}, J.~M., {D'Amico}, N., {Camilo}, F., {Joshi},
  B.~C., \& {Freire}, P.~C.~C. 2005, ApJ, 621, L49

\bibitem[{Manchester {et~al.}(2001)Manchester, Lyne, Camilo, Bell, Kaspi,
  D'Amico, McKay, Crawford, Stairs, Possenti, Morris, \& Sheppard}]{mlc+01}
Manchester, R.~N., Lyne, A.~G., Camilo, F., Bell, J.~F., Kaspi, V.~M., D'Amico,
  N., McKay, N. P.~F., Crawford, F., Stairs, I.~H., Possenti, A., Morris,
  D.~J., \& Sheppard, D.~C. 2001, MNRAS, 328, 17

\bibitem[{Manchester {et~al.}(1996)Manchester, Lyne, D'Amico, Bailes, Johnston,
  Lorimer, Harrison, Nicastro, \& Bell}]{mld+96}
Manchester, R.~N., Lyne, A.~G., D'Amico, N., Bailes, M., Johnston, S., Lorimer,
  D.~R., Harrison, P.~A., Nicastro, L., \& Bell, J.~F. 1996, MNRAS, 279, 1235

\bibitem[{{McLaughlin} {et~al.}(2004){McLaughlin}, {Lyne}, {Lorimer},
  {Possenti}, {Manchester}, {Camilo}, {Stairs}, {Kramer}, {Burgay}, {D'Amico},
  {Freire}, {Joshi}, \& {Bhat}}]{mll+04}
{McLaughlin}, M.~A., {Lyne}, A.~G., {Lorimer}, D.~R., {Possenti}, A.,
  {Manchester}, R.~N., {Camilo}, F., {Stairs}, I.~H., {Kramer}, M., {Burgay},
  M., {D'Amico}, N., {Freire}, P.~C.~C., {Joshi}, B.~C., \& {Bhat}, N.~D.~R.
  2004, ApJ, 616, L131

\bibitem[{Narayan \& Vivekanand(1983)}]{nv83}
Narayan, R. \& Vivekanand, M. 1983, A\&A, 122, 45

\bibitem[{Navarro {et~al.}(1997)Navarro, Manchester, Sandhu, Kulkarni, \&
  Bailes}]{nms+97}
Navarro, J., Manchester, R.~N., Sandhu, J.~S., Kulkarni, S.~R., \& Bailes, M.
  1997, ApJ, 486, 1019

\bibitem[{{Ng} \& {Romani}(2008)}]{nr08}
{Ng}, C.-Y. \& {Romani}, R.~W. 2008, ApJ, 673, 411

\bibitem[{{Ord} {et~al.}(2002){Ord}, {Bailes}, \& {van Straten}}]{obv02}
{Ord}, S.~M., {Bailes}, M., \& {van Straten}, W. 2002, ApJ, 574, L75

\bibitem[{Peters(1964)}]{pet64}
Peters, P.~C. 1964, Phys. Rev., 136, 1224

\bibitem[{{Piran} \& {Shaviv}(2005)}]{ps05}
{Piran}, T. \& {Shaviv}, N.~J. 2005, Phys. Rev. Lett., 94, 051102

\bibitem[{{Podsiadlowski} {et~al.}(2005){Podsiadlowski}, {Dewi}, {Lesaffre},
  {Miller}, {Newton}, \& {Stone}}]{pdl+05}
{Podsiadlowski}, P., {Dewi}, J.~D.~M., {Lesaffre}, P., {Miller}, J.~C.,
  {Newton}, W.~G., \& {Stone}, J.~R. 2005, MNRAS, 361, 1243

\bibitem[{Radhakrishnan \& Cooke(1969)}]{rc69a}
Radhakrishnan, V. \& Cooke, D.~J. 1969, Astrophys. Lett., 3, 225

\bibitem[{{Ramachandran} {et~al.}(2004){Ramachandran}, {Backer}, {Rankin}, M.,
  \& E.}]{rbr+04}
{Ramachandran}, R., {Backer}, D.~C., {Rankin}, J.~M., M., W.~J., \& E., D.~K.
  2004, ApJ, 606, 1167

\bibitem[{Ransom {et~al.}(2004)Ransom, Kaspi, Ramachandran, Demorest, Backer,
  Pfahl, Ghigo, \& Kaplan}]{rkr+04}
Ransom, S.~M., Kaspi, V.~M., Ramachandran, R., Demorest, P., Backer, D.~C.,
  Pfahl, E.~D., Ghigo, F.~D., \& Kaplan, D.~L. 2004, ApJ, 609, L71

\bibitem[{Spruit \& Phinney(1998)}]{sp98}
Spruit, H. \& Phinney, E.~S. 1998, Nature, 393, 139

\bibitem[{{Stairs} {et~al.}(2004){Stairs}, {Thorsett}, \&
  {Arzoumanian}}]{sta04}
{Stairs}, I.~H., {Thorsett}, S.~E., \& {Arzoumanian}, Z. 2004, Phys. Rev.
  Lett., 93, 141101

\bibitem[{{Stairs} {et~al.}(2006){Stairs}, {Thorsett}, {Dewey}, {Kramer}, \&
  {McPhee}}]{std+06}
{Stairs}, I.~H., {Thorsett}, S.~E., {Dewey}, R.~J., {Kramer}, M., \& {McPhee},
  C.~A. 2006, MNRAS, 373, L50

\bibitem[{Standish(1998)}]{sta98b}
Standish, E.~M. 1998, JPL Planetary and Lunar Ephemerides, DE405/LE405, Memo
  IOM 312.F-98-048 (Pasadena: JPL),
  http://ssd.jpl.nasa.gov/iau-comm4/de405iom/de405iom.pdf

\bibitem[{Staveley-Smith {et~al.}(1996)Staveley-Smith, Wilson, Bird, Disney,
  Ekers, Freeman, Haynes, Sinclair, Vaile, Webster, \& Wright}]{swb+96}
Staveley-Smith, L., Wilson, W.~E., Bird, T.~S., Disney, M.~J., Ekers, R.~D.,
  Freeman, K.~C., Haynes, R.~F., Sinclair, M.~W., Vaile, R.~A., Webster, R.~L.,
  \& Wright, A.~E. 1996, PASA, 13, 243

\bibitem[{{Tauris} \& {Sennels}(2000)}]{ts00a}
{Tauris}, T.~M. \& {Sennels}, T. 2000, A\&A, 355, 236

\bibitem[{Thorsett \& Chakrabarty(1999)}]{tc99}
Thorsett, S.~E. \& Chakrabarty, D. 1999, ApJ, 512, 288

\bibitem[{Thorsett {et~al.}(2005)Thorsett, Dewey, \& Stairs}]{tds05}
Thorsett, S.~E., Dewey, R.~J., \& Stairs, I.~H. 2005, ApJ, 619, 1036

\bibitem[{{van Straten}(2004)}]{vs04}
{van Straten}, W. 2004, ApJS, 152, 129, preprint, astro-ph/0401536

\bibitem[{{van Straten} {et~al.}(2009){van Straten}, {Manchester}, {Johnston},
  \& {Reynolds}}]{vmjr09}
{van Straten}, W., {Manchester}, R.~N., {Johnston}, S., \& {Reynolds}, J. 2009,
  PASA, in press

\bibitem[{{Wang} {et~al.}(2007){Wang}, {Manchester}, \& {Johnston}}]{wmj07}
{Wang}, N., {Manchester}, R.~N., \& {Johnston}, S. 2007, MNRAS, 377, 1383

\bibitem[{Wang {et~al.}(2000)Wang, Manchester, Pace, Bailes, Kaspi, Stappers,
  \& Lyne}]{wmp+00}
Wang, N., Manchester, R.~N., Pace, R., Bailes, M., Kaspi, V.~M., Stappers,
  B.~W., \& Lyne, A.~G. 2000, MNRAS, 317, 843

\bibitem[{Weisberg {et~al.}(1989)Weisberg, Romani, \& Taylor}]{wrt89}
Weisberg, J.~M., Romani, R.~W., \& Taylor, J.~H. 1989, ApJ, 347, 1030

\bibitem[{{Weisberg} \& {Taylor}(2002)}]{wt02}
{Weisberg}, J.~M. \& {Taylor}, J.~H. 2002, ApJ, 576, 942

\bibitem[{{Weisberg} \& {Taylor}(2005)}]{wt05}
{Weisberg}, J.~M. \& {Taylor}, J.~H. 2005, in {Binary Radio Pulsars}, ed.
  F.~Rasio \& I.~H. Stairs (San Francisco: Astronomical Society of the
  Pacific), 25--31

\bibitem[{Wex {et~al.}(2000)Wex, Kalogera, \& Kramer}]{wkk00}
Wex, N., Kalogera, V., \& Kramer, M. 2000, ApJ, 528, 401

\bibitem[{Willems {et~al.}(2004)Willems, Kalogera, \& Henninger}]{wkh04}
Willems, B., Kalogera, V., \& Henninger, M. 2004, ApJ, 616, 414

\bibitem[{{Willems} {et~al.}(2006){Willems}, {Kaplan}, {Fragos}, {Kalogera}, \&
  {Belczynski}}]{wkf+06}
{Willems}, B., {Kaplan}, J., {Fragos}, T., {Kalogera}, V., \& {Belczynski}, K.
  2006, Phys. Rev. D, 74, 043003

\end{thebibliography}

\begin{deluxetable}{cccccccc}
\tablecaption{Observations of PSR J1141$-$6545\label{tb:obs}}
\tabletypesize{\scriptsize}
\tablehead{
\colhead{Mean} & \colhead{Mean} & \colhead{MJD Range} & \colhead{Receiver} &
\colhead{Backend} & \colhead{Ctr. Freq.} & \colhead{Nr of} & \colhead{Int. Time} \\
\colhead{Date} & \colhead{MJD} & & & \colhead{System} & \colhead{(MHz)} & 
\colhead{Observations} & \colhead{(h)}
}
\startdata
1999/08/21 & 51411.7 & 51407--51413 & MB   & AFB & 1390 & 11 & 1.72 \\
1999/10/02 & 51452.9 & 51451--51454 & MB   & AFB & 1390 & 8  & 1.23 \\
1999/12/01 & 51513.5 & 51498--51529 & MB   & AFB & 1390 & 4  & 0.66 \\
2000/02/04 & 51577.7 & 51554--51632 & MB   & AFB & 1390 & 12 & 2.10 \\
2000/06/27 & 51721.6 & 51710--51754 & MB   & AFB & 1390 & 4  & 0.99 \\
2000/11/23 & 51851.6 & 51841--51940 & MB   & AFB & 1390 & 5  & 0.80 \\
2001/04/24 & 52022.9 & 51969--52115 & MB   & AFB & 1390 & 8  & 0.78 \\
2001/10/20 & 52201.9 & 52132--52251 & MB   & AFB & 1390 & 6  & 0.60 \\
2002/03/28 & 52361.0 & 52305--52428 & MB   & AFB & 1390 & 6  & 0.60 \\
2002/07/30 & 52485.0 & 52459--52507 & MB   & AFB & 1390 & 8  & 1.14 \\
2002/11/21 & 52598.9 & 52571--52624 & MB   & AFB & 1390 & 6  & 0.60 \\
2003/03/17 & 52715.4 & 52660--52770 & MB   & AFB & 1390 & 4  & 0.40 \\
2003/11/19 & 52961.6 & 52914--53004 & MB/H-OH & AFB & 1390 & 6  & 0.52 \\
2004/04/07 & 53102.5 & 53102--53102 & H-OH & WBC & 1375 & 1  & 1.00 \\
2004/04/23 & 53118.0 & 53103--53146 & H-OH & AFB & 1390 & 3  & 0.30 \\
2004/07/04 & 53190.4 & 53183--53194 & H-OH & AFB & 1390 & 5  & 3.20 \\
2004/07/07 & 53193.3 & 53193--53194 & H-OH & WBC & 1375 & 8  & 8.00 \\
2004/08/31 & 53248.2 & 53223--53282 & H-OH/MB & AFB & 1390 & 3  & 0.30 \\
2004/10/31 & 53309.6 & 53309--53311 & MB   & WBC & 1433 & 12 & 11.76 \\
2004/11/01 & 53310.2 & 53306--53311 & MB   & AFB & 1390 & 8  & 3.85 \\
2005/01/02 & 53372.1 & 53371--53372 & MB   & WBC & 1433 & 3  & 3.00 \\
2005/01/02 & 53372.5 & 53371--53372 & MB   & AFB & 1390 & 6  & 5.70 \\
2005/04/23 & 53482.9 & 53482--53484 & MB   & AFB & 1390 & 9  & 8.55 \\
2005/06/05 & 53526.1 & 53526--53526 & MB   & WBC & 1433 & 2  & 2.00 \\
2005/08/04 & 53586.1 & 53586--53586 & MB   & WBC & 1433 & 12 & 7.70 \\
2005/08/13 & 53594.6 & 53522--53620 & MB   & AFB & 1390 & 4  & 2.16 \\
2005/09/05 & 53618.7 & 53617--53620 & MB   & PDFB1 & 1433 & 3  & 2.05 \\
2005/12/27 & 53730.8 & 53714--53741 & MB   & AFB & 1390 & 3  & 1.55 \\
2006/01/05 & 53740.8 & 53735--53745 & MB   & PDFB1 & 1433 & 2  & 1.95 \\
2006/02/04 & 53770.5 & 53770--53770 & MB   & PDFB1 & 1433 & 2  & 1.52 \\
2006/03/03 & 53797.7 & 53797--53797 & MB   & PDFB1 & 1369 & 2  & 1.67 \\
2006/04/09 & 53834.5 & 53834--53834 & MB   & PDFB1 & 1369 & 2  & 1.67 \\
2006/05/10 & 53865.5 & 53865--53865 & MB   & PDFB1 & 1369 & 2  & 1.67 \\
2006/05/28 & 53883.5 & 53883--53883 & MB   & PDFB1 & 1369 & 2  & 1.67 \\
2006/07/10 & 53926.7 & 53921--53932 & MB   & PDFB1 & 1369 & 2  & 1.37 \\
2006/08/14 & 53961.3 & 53961--53961 & MB   & PDFB1 & 1369 & 2  & 1.67 \\
2006/09/27 & 54005.5 & 53995--54015 & MB   & PDFB1 & 1369 & 4  & 2.83 \\
2007/01/19 & 54119.8 & 54119--54120 & H-OH & PDFB1 & 1433 & 4  & 3.00 \\
2007/02/09 & 54140.7 & 54137--54147 & H-OH & PDFB1 & 1433 & 3  & 1.63 \\
2007/05/03 & 54223.5 & 54223--54223 & H-OH & PDFB1 & 1433 & 4  & 3.33 \\
2007/06/06 & 54257.4 & 54257--54257 & MB   & PDFB1 & 1369 & 2  & 1.67 \\
2007/07/18 & 54299.2 & 54299--54299 & MB   & PDFB1 & 1369 & 2  & 2.13 \\
2007/08/04 & 54316.0 & 54315--54316 & MB   & PDFB1 & 1369 & 2  & 2.13 \\
2007/11/05 & 54409.1 & 54409--54409 & MB   & PDFB1 & 1369 & 2  & 2.13 \\
2008/01/24 & 54489.6 & 54489--54489 & MB   & PDFB1 & 1369 & 1  & 1.06 \\
2008/02/24 & 54509.6 & 54509--54509 & MB   & PDFB1 & 1369 & 1  & 1.06 \\
2008/08/03 & 54681.2 & 54681--54681 & MB   & PDFB3 & 1369 & 2  & 2.11 \\
2008/11/14 & 54784.8 & 54784--54784 & MB   & PDFB3 & 1369 & 2  & 2.13 \\
\enddata
\end{deluxetable}

\begin{deluxetable}{ll}
\tablecaption{PSR J1141$-$6545 glitch parameters\label{tb:glt}}
\tablehead{
\colhead{Parameter} & \colhead{Value}}
\startdata
Glitch epoch (MJD)  &  $54277 \pm 20$ \\
$\Delta\nu_g/\nu$   &  $(5.890 \pm 0.006)\times 10^{-7}$ \\
$\Delta\dot\nu_g/\dot\nu$ & $(5.0 \pm 0.9)\times 10^{-3}$ \\
$Q$                 &  $0.0040 \pm 0.0007$  \\
$\tau_d$ (d)        &  $495 \pm 140$ \\
Data span (MJD)     &  53834 -- 54785 \\
Number of ToAs      &  254 \\
Rms residual ($\mu$s) & 220 
\enddata
\end{deluxetable}

\begin{deluxetable}{ll}
\tablecaption{PSR J1141$-$6545 post-glitch timing parameters\label{tb:psrpar}}
\tablehead{
\colhead{Parameter} & \colhead{Value\tablenotemark{a}}}
\startdata
R.A. (J2000)  &  11$^{\rm h}$ 41$^{\rm m}$ $07\fs0140$\tablenotemark{b} \\
Dec. (J2000)  &  $-65\degr$ $45\arcmin$ $19\farcs1131$\tablenotemark{b} \\
Pulse Frequency ($\nu$) (Hz) & 2.538723048486(4)\\
Pulse Frequency time-deriv. ($\dot\nu$) (s$^{-2}$) & $-2.77615(8) \times 10^{-14}$ \\
Epoch (MJD)  & 54637.00 \\
Dispersion Measure (cm$^{-3}$ pc) & 116.080\tablenotemark{b} \\
Binary Period (d) & 0.1976509593\tablenotemark{b} \\
Binary Period time-derivative & $-4.3 \times 10^{-13}$\tablenotemark{b} \\
Orbit semi-major axis (s) & 1.858922\tablenotemark{b} \\
Eccentricity  & 0.171884\tablenotemark{b} \\
Periastron time (MJD) & 51369.8545515\tablenotemark{b} \\
Longitude of periastron ($\degr$) & 42.4561\tablenotemark{b} \\
Longitude time-derivative ($\degr$ yr$^{-1}$) & 5.3096\tablenotemark{b} \\
Grav. redshift/time dilation ($\gamma$) (s) & 0.000773\tablenotemark{b} \\
Data span (MJD) & 54435 -- 54785 \\
Rms timing residual ($\mu$s) & 56
\enddata
\tablenotetext{a}{Uncertainties in the last quoted digit are given in parentheses.}
\tablenotetext{b}{From \citet{bbv08}.}
\end{deluxetable}

\begin{deluxetable}{lcccccc}
\tablecaption{Central phases for the profile gaussian components\label{tb:comp_phs}}
\tablehead{
\colhead{Component} & \colhead{1}& \colhead{2} & \colhead{3} &
\colhead{4} & \colhead{5} & \colhead{6} }
\startdata
Central phase & $-0.00892$ & 0.00264 & 0.00642 & 0.00822 & 0.01325 &
0.01667 \\
\enddata
\end{deluxetable}

\begin{deluxetable}{lc}
\tablecaption{Model parameters from the RVM fit to the observed
  position-angle variations \label{tb:model}}
\tablehead{
\colhead{Parameter} & \colhead{Value\tablenotemark{a}}}
\startdata
Assumed parameters: & \\
Orbit inclination $i$         & $73\degr$ \\
Precession rate $\Omega_p$      & $1\fdg36$ yr$^{-1}$ \\
RVM reference phase $\phi_0$  & 0.0 \\
Reference time for precession & MJD 53000.0 \\
Derived parameters: & \\
Spin-orbit misalignment angle $\delta$  & $93\degr$ ($-9\degr$,$+16\degr$) \\
Precession angle $\Phi_0$     & $175\fdg5$ ($-2\fdg8$,$+1\fdg8$) \\ 
Offset in PA  $\Delta\psi_0$ & $-118\fdg7 \pm 0\fdg2$ \\ 
Magnetic inclination $\alpha$ & $160\degr$ ($-16\degr$,$+8\degr$) \\
\enddata
\tablenotetext{a}{Most probable value and 68\% confidence limits.}
\end{deluxetable}

\begin{deluxetable}{lcccc}
\tablecaption{Median values and confidence limits for parameters
    derived from the constrained model distributions\label{tb:pdf}}
\tablehead{ \colhead{Parameter} & Vel. Disp. (km s$^{-1}$) &
  \colhead{Median} & \colhead{68\% limits} & \colhead{95\% limits} } 
\startdata
Progenitor mass (M$_\odot$) & 50  & 1.80 & 1.54 -- 2.10 & 1.42 -- 2.48  \\
                           & 100  & 2.24 & 1.58 -- 3.06 & 1.42 -- 4.00  \\
Kick velocity (km s$^{-1}$) & 50  & 116 & 86 -- 164 & 56 -- 240  \\
                            & 100  & 258 & 150 -- 434 & 94 -- 452  \\
Spin-orbit misalignment angle ($\degr$)& 50  & 116 & 101 -- 131 & 92 -- 149 \\
                            & 100  & 106 & 94 -- 120 & 87 -- 135 \\
\enddata
\end{deluxetable}

\begin{figure*}
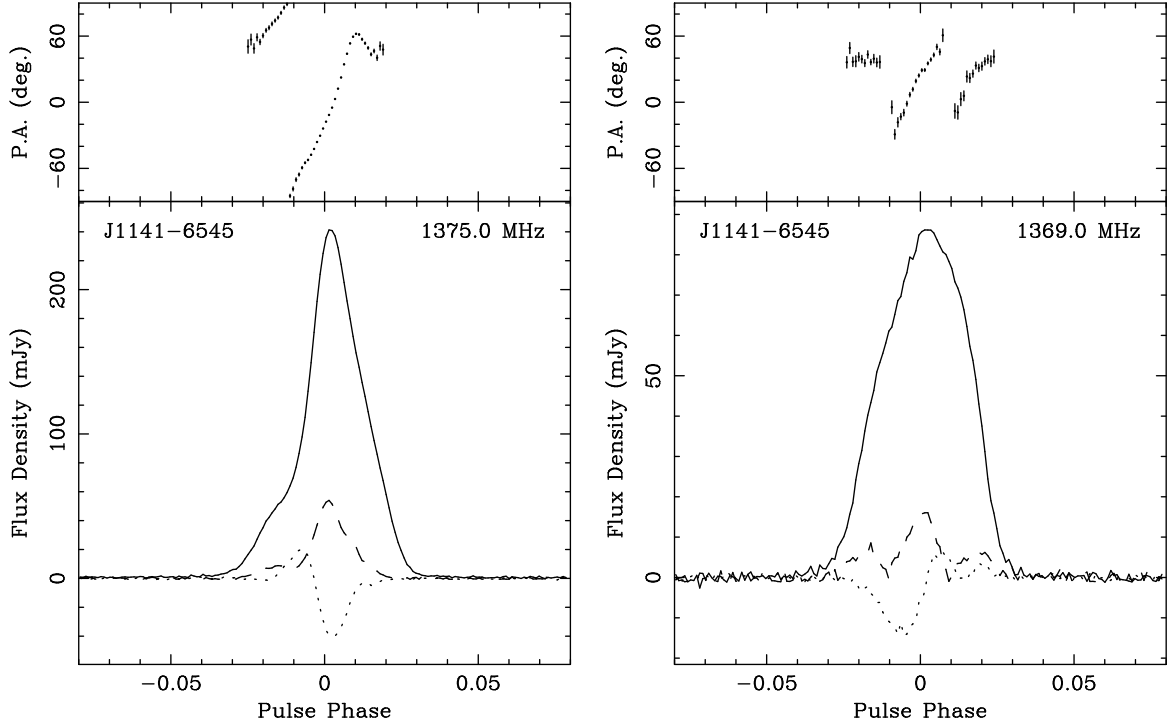

\begin{center} 
\begin{tabular}{cc}
\mbox{\includegraphics[width=75mm]{w040707_poln.ps}} &
\mbox{\includegraphics[width=75mm]{a070803_poln.ps}} 
\end{tabular}
\caption{Mean pulse profiles and polarization parameters for PSR
  J1141$-$6545 at two epochs, 2004 July 07 (MJD 53193) (left) and 2007
  August 04 (MJD 54316) (right). In the lower panel of each plot the
  solid line is the total intensity (Stokes $I$), the dashed line is
  the linearly polarized intensity $L = (U^2 + Q^2)^{1/2}$ and the
  dotted line is Stokes $V = I_{LH}-I_{RH}$. The position angle of the
  linearly polarized component is shown in the upper panel of each
  plot. Note that postion angles are defined according to the IAU
  convention; they are absolute and apply to the frequency marked on
  each plot.}\label{fg:poln}
\end{center}
\end{figure*}

\clearpage

\end{document}